\documentclass[final]{elsarticle}

\usepackage{natbib}    
\usepackage{graphicx}  
\usepackage{array}     
\usepackage{color}     

\begin{document}

\begin{frontmatter}
\title{The GENIE Neutrino Monte Carlo Generator}


\author[RAL]        {C.Andreopoulos}
\author[Pitt]       {A.Bell}
\author[Pitt]       {D.Bhattacharya}
\author[LAquila]    {F.Cavanna}
\author[IC]         {J.Dobson}
\author[Pitt]       {S.Dytman}
\author[Tufts]      {H.Gallagher}
\author[IC]         {P.Guzowski} 
\author[FNAL]       {R.Hatcher} 
\author[Tufts]      {P.Kehayias}
\author[ETHZ,IPHC]  {A.Meregaglia}
\author[Pitt]       {D.Naples}
\author[RAL]        {G.Pearce}
\author[ETHZ]       {A.Rubbia}
\author[Durham]     {M.Whalley}
\author[Stanford]   {T.Yang}

\address[RAL]      {Rutherford Appleton Laboratory, STFC, Oxfordshire OX11 0QX, UK}
\address[Pitt]     {Physics Dept., Pittsburgh University, Pittsburgh PA 15260, USA}
\address[LAquila]  {Physics Dept., L'Aquila University, 67100 L'Aquila, Italy}
\address[IC]       {Physics Dept., Imperial College London, Blackett Lab., London SW7 2BW, UK}
\address[Tufts]    {Physics Dept., Tufts University, Medford, MA, 02155, USA}
\address[FNAL]     {Fermi National Accelerator Laboratory, Batavia, Illinois 60510, USA}
\address[ETHZ]     {Physics Dept., ETH Zurich, CH-8093 Zurich, Switzerland}
\address[IPHC]     {IPHC - Strasbourg, F-67037 Strasbourg Cedex 2, France}
\address[Durham]   {Physics Dept., Durham University, Durham DH1 3LE, UK}
\address[Stanford] {Physics Dept., Stanford University, Stanford, CA 94309, USA}


\begin{abstract}

GENIE \cite{GENIE} is a new neutrino event generator for the experimental 
neutrino physics community.  The goal of the project is to develop a 
`canonical' neutrino interaction physics Monte Carlo 
whose validity extends to all nuclear targets and neutrino flavors from MeV to PeV energy scales.
Currently, emphasis is on the few-GeV energy range, the challenging boundary between the 
non-perturbative and perturbative regimes, which is relevant for the current and near future 
long-baseline precision neutrino experiments using accelerator-made beams.
The design of the package addresses many challenges unique to neutrino 
simulations and supports the full life-cycle of
simulation and generator-related analysis tasks.   

GENIE is a large-scale software system,
consisting of $\sim$120,000 lines of C++ code, featuring a modern object-oriented 
design and extensively validated physics content.  
The first official physics release of GENIE was made available in August 2007, and at the 
time of the writing of this article, the latest available version was v2.4.4.  

\end{abstract}


\end{frontmatter}



\section{Introduction}

Over the last few years, throughout the field of high energy physics (HEP), we have witnessed 
an enormous effort committed to migrating many popular procedural Monte Carlo Generators into 
their C++ equivalents designed using object-oriented methodologies.
Well-known examples are the
GEANT \cite{Agostinelli:2002hh}, 
HERWIG \cite{Bahr:2008pv} and 
PYTHIA \cite{Sjostrand:2007gs} Monte Carlo Generators.
This reflects a radical change in our approach to scientific computing.
Along with the eternal requirement that the modeled physics be correct and extensively 
validated with external data, the evolving nature of computing in HEP has introduced
new requirements.   
These requirements relate to the way large HEP software systems are developed and maintained,
by wide geographically-spread collaborations over a typical time-span of $\sim$25 years
during which they will undergo many (initially unforeseen) extensions and modifications to 
accommodate new needs. This puts a stress on software qualities such as re-usability, maintainability, 
robustness, modularity and extensibility.  Software engineering provides many well proven techniques 
to address these requirements and thereby improves the quality and lifetime of HEP software.
In neutrino physics, the requirements for a new physics generator are more challenging 
for three reasons:  the lack of a `canonical' procedural generator, theoretical and phenomenological 
challenges in modeling few-GeV neutrino interactions, and the rapidly evolving experimental and
theoretical landscape.   

Neutrinos come from many sources and a variety of experiments have
been mounted to measure their properties.  These experiments have
complicated detectors composed of many elements and the neutrinos
have many flavors and a wide energy spectrum (from $\sim$ 1 MeV
to $\sim$ 1 PeV).  Our long-term goal
is for GENIE to become a `canonical' neutrino event generator 
with wide applicability.
The origins of the code come from the Soudan experiment \cite{Gallagher:1998qg} and recent
application has been primarily to MINOS \cite{Michael:2006rx}.  Thus, emphasis has been given
to the few-GeV energy range, the challenging boundary between the
non-perturbative and perturbative regimes.  These are relevant for 
current and near future long-baseline precision neutrino oscillation
experiments using accelerator-made beams, one of the focuses
of high energy physics.  GENIE development over the next five years will be driven 
by the upcoming generation of accelerator experiments including T2K \cite{Wark:2008zz}, 
NoVA \cite{Ayres:2004js}, Minerva \cite{Schulte:2006db}
MicroBooNE \cite{Soderberg:2009rz} and ArgoNEUT \cite{Soderberg:2009qt}.
These developments are well underway and the code is being used successfully in each of these
experiments.  The present version provides comprehensive neutrino
interaction modelling in the energy from $\sim$100 MeV to a few hundred GeV.
Results can be obtained and will be qualitatively correct for any nuclear target.

GENIE\footnote {
 \em GENIE \em stands for
 \em\underline{G}\em enerates
 \em\underline{E}\em vents for
 \em\underline{N}\em eutrino
 \em\underline{I}\em nteraction
 \em\underline{E}\em xperiments
} is a ROOT-based \cite{Brun:1997pa} Neutrino MC Generator.
It was designed using object-oriented methodologies and developed entirely in C++ over a 
period of more than three years, from 2004 to 2007. 
Its first official physics release (v2.0.0) was made available on August 2007 and, at the time 
of writing this article, the latest available version was v2.4.4.  It also describes
v2.6.0 which will be released shortly.
GENIE has already been adopted by the majority of neutrino experiments, including 
those using the JPARC and NuMI neutrino beamlines, and will be an important physics tool
for the exploitation of the world accelerator neutrino program.

The project is supported by a group of physicists from all major neutrino experiments
operating in this energy range,  establishing GENIE as a major HEP event generator collaboration.
Many members of the GENIE collaboration have extensive experience in developing 
and maintaining the legacy Monte Carlo Generators that GENIE strives to replace,
which guarantees knowledge exchange and continuation.  The default set of physics models in GENIE have 
adiabatically evolved from those in the NEUGEN \cite{Gallagher:2002sf} package, 
which was used as the event generator by numerous experiments over the past decade.  

This article will discuss the paradigm shift brought about by GENIE in neutrino physics simulations.
In Sec. \ref{sec:bridge} we describe the unique challenges facing neutrino simulations in more detail.  
Section \ref{sec:physics}
gives a brief overview of the physics models available in GENIE.  Section \ref{future} gives
a brief discussion of upgrades in progress.
Section \ref{sec:design} describes
the object-oriented design of GENIE. 
Section \ref{sec:apps} describes the GENIE applications and utilities available 
for simulation and analysis tasks.  Section \ref{sec:collab} describes the structure of the GENIE 
collaboration and 
Section \ref{sec:avail} describes code availability, 
distribution, supported platforms, external dependencies, releases, and license.  

\section{Neutrino Interaction Simulation:  Challenges and Significance}
\label{sec:bridge}

Neutrinos have played an important role in particle physics since their discovery
half a century ago.  They have been used to elucidate the structure of the electroweak
symmetry groups, to illuminate the quark nature of hadrons, and to confirm our models of
astrophysical phenomena.   With the discovery of neutrino oscillations
using atmospheric, solar, accelerator, and reactor neutrinos, these elusive particles 
now take center stage as the objects of study themselves.  Precision measurements of the lepton mixing matrix, 
the search for lepton charge-parity (CP) violation, and the determination of the neutrino masses and 
 hierarchy will be major efforts in HEP for several decades. 
The cost of this next generation of experiments will be
significant, typically tens to hundreds of millions of dollars.
A comprehensive, thoroughly tested neutrino event generator package
plays an important role in the design and execution of these experiments, since this tool is used 
to evaluate the feasibility of proposed projects and estimate their physics impact, make
decisions about detector design and optimization, analyze the collected data samples, and
evaluate systematic errors. With the advent of high-intensity neutrino beams
from proton colliders, we have entered a new era of
high-statistics, precision neutrino experiments which
will require a new level of accuracy
in our knowledge, and simulation,  of neutrino interaction physics \cite{Harris:2004iq}.

While object-oriented physics generators in other fields of high energy physics were
evolved from well established legacy systems, in neutrino physics no such `canonical' MC exists.
Until quite recently, most neutrino experiments developed their own neutrino event generators.
This was due partly to the wide variety of energies, nuclear targets, detectors, and physics topics being
simulated.   
Without doubt these generators, 
the most commonly used of which have been GENEVE \cite{Cavanna:2002se}, NEUT \cite{Hayato:2002sd},
NeuGEN \cite{Gallagher:2002sf}, NUANCE \cite{Casper:2002sd} and NUX \cite{Rubbia:2001},
played an important role in the design and exploration of the previous and current generation
of accelerator neutrino experiments.
Tuning on the basis of unpublished data from each group's own experiment has not been unusual
making it virtually impossible to perform a global, independent evaluation for the state-of-the-art
in neutrino interaction physics simulations.
Moreover, limited manpower and the fragility of the overextended software architectures 
meant that many of these legacy physics generators were not keeping
up with the latest theoretical ideas and experimental measurements.
A more recent development in the field has been the direct involvement of theory groups in the 
development of neutrino event generators, such as the NuWRO \cite{Juszczak:2005zs} 
and GiBUU \cite{Leitner:2006ww} packages, and the inclusion of neutrino scattering in 
the long-established FLUKA hadron scattering package \cite{Fasso:2003xz}.

Simulating neutrino interactions in the energy range of interest to current and near-future experiments poses significant challenges. 
This broad energy range bridges the perturbative
and non-perturbative pictures of the nucleon and a variety of scattering mechanisms are important.
In many areas, including elementary cross sections, hadronization models, and nuclear physics, one is required
to piece together models with different ranges of validity in order to generate events over all of the available phase space. 
This inevitably introduces challenges in merging and tuning models, making sure that double counting and discontinuities 
are avoided.  In addition there are kinematic regimes which are outside the stated range of validity of all available models, 
in which case we are left with the challenge of developing our own models or deciding which model best extrapolates into 
this region.  
An additional fundamental problem in this energy range is a lack of data.  Most simulations have been tuned to bubble chamber
data taken in the 70's and 80's.  Because of the limited size of the data samples (important exclusive channels
might only contain a handful of events), and the limited coverage 
in the space of ($\nu/\overline{\nu}, E_\nu$, A), substantial uncertainties exist in numerous aspects of
the simulations.  

The use cases for GENIE are also informed by the experiences of the developers and users
of the previous generation of procedural codes.  
Dealing with these substantial model uncertainties has been an important analysis challenge for many recent experiments. 
The impact of these uncertainties on physics analyses have been evaluated in detailed systematics studies and in some 
cases the models have been fit directly to experimental data to reduce systematics.  These `downstream' simulation-related
studies can often be among the most challenging and time-consuming in an analysis.    

To see the difficulties facing the current generation of neutrino experiments, one can
look no further than the K2K and MiniBooNE experiments.  Both of these experiments have
measured a substantially different Q$^2$ distribution for their quasielastic-like events
when compared with their simulations, which involve a standard Fermi Gas model nuclear model
\cite{Ishida:2002xd,miniBoone:2007ru}.
The disagreement between nominal Monte Carlo and data is quite large - in the lowest Q$^2$
bin of MiniBooNE the deficit in the data is around 30\% \cite{miniBoone:2007ru}.  It seems likely that the discrepancies seen by both experiments have a common origin.
However the two experiments have been able to obtain internal consistency with very different
model changes - the K2K experiment does this by eliminating the charged current (CC) coherent contribution in the
Monte Carlo \cite{Sanchez:2006hp} and the MiniBooNE experiment does this by modifying certain parameters in their
Fermi Gas model \cite{miniBoone:2007ru}.  Another example of the rapidly evolving nature of this field  
is the recently reported excess of low energy electron-like events by the MiniBooNE collaboration \cite{AguilarArevalo:2008rc}.  
These discrepancies have generated significant new theoretical 
work on these topics over the past several years \cite{Paschos:2005km,
Singh:2006bm,AlvarezRuso:2007tt,Bodek:2007wb,Harvey:2007rd,Buss:2007ar,Benhar:2005dj,Amaro:2006if}.   
The situation is bound to become even more interesting, and complicated, in the coming decade, 
as new high-statistics experiments begin taking data in this energy range. 
Designing a software system that can be responsive to this rapidly 
evolving experimental and theoretical landscape is a major challenge. 

In this paper we will describe the ways in which the GENIE neutrino event generator addresses these 
challenges. These solutions rely heavily on the power of modern software engineering, particularly 
the extensibility, modularity, and flexibility of object oriented design, as well as the combined
expertise and experience of the collaboration with previous procedural codes.

\section{Neutrino Interaction Physics models in GENIE}
\label{sec:physics}

The set of physics models used in GENIE incorporates the dominant scattering mechanisms
from several MeV to several hundred GeV and are appropriate for any neutrino
flavor and target type.  Over this energy range, many different physical 
processes are important.
These physics models can be 
broadly categorized into nuclear physics models, cross section models, and hadronization models.  

The neutrino-nucleus interaction involves a large variety of processes, all of which
must be modeled to get an accurate description of the experimental signature
of any detector and its many components. Since most theoretical models 
describe a small subset of these processes, GENIE must include many models.
The broad energy range and the many nuclei to be covered force choosing 
models that have very broad applicability.  

The particle in the nucleus with which the neutrino interacts depends strongly on
the energy.  At high energies ($E_\nu >$10 GeV) neutrinos interact with a single quark 
inside a nucleon (neutron or
proton); the code must model this interaction and the distribution of the residual quarks.  
At lower energies, the struck particles are neutrons and protons.  The neutrino tends to 
strike a single nucleon (impulse approximation) which is affected by the nuclear medium
in which it resides.  
In the high energy regime, the large body of neutrino-nucleon data is sufficient for
development of a full model.  At lower energies, neutrino-nucleon data has been
used for the basic process and nuclear models developed for other probes (especially
electrons) are adopted.


The 2 recent major developments have been in the transition region
and the final state interactions (FSI) model.  
In physics model development for GENIE we have been forced 
to pay particular attention to this `transition region', as for few-GeV
experiments it dominates the event rate.   
In particular the boundaries between regions of validity of 
different models need to be treated with care in order to 
avoid theoretical inconsistencies, 
discontinuities in generated distributions, and double-counting.  
Treatment of FSI involves many aspects of nuclear physics and strong interactions.
Many events where particles are produced by the neutrinos have their 
topologies and kinematics altered.  There are many effects to include
and some dispute about the right techniques to use.  Therefore, FSI
treatment is one of the largest differences among models of the neutrino-nucleus
interaction.

In this brief section we will describe the models available in GENIE 
and  the ways in which we combine models to cover
regions of phase space where clear theoretical or empirical guidance is lacking. 




\subsection{Nuclear Physics Model}

The importance of the nuclear model depends strongly on the kinematics of the reaction. 
Nuclear physics plays a large role in every aspect of neutrino scattering simulations 
at few-GeV energies and introduces coupling between several aspects of the simulation.
The relativistic Fermi gas (RFG) nuclear model is used for all processes.  
GENIE uses the version of Bodek and Ritchie which has been modified 
to incorporate short range nucleon-nucleon correlations \cite{Bodek:1981wr}.  This is simple,
yet applicable across a broad range of target atoms and neutrino energies.  The best 
tests of the RFG model come from electron scattering
experiments \cite{Moniz:1971mt}. At high energies, the nuclear model requires
broad features due to shadowing and similar effects.  
At the lower end of the GENIE energy range, the impulse
approximation works very well and the RFG is often successful.
The nuclear medium gives the struck nucleon a momentum and average binding energy
which have been determined in electron scattering experiments.
Mass densities are taken from review articles \cite{DeJager:1987qc}.  For $A<$20, the
modified Gaussian density parameterization is used.  For heavier nuclei, the
2-parameter Woods-Saxon density function is used.  Thus, the model can be used
for {\em all} nuclei.  Presently, fit parameters for selected nuclei are used
with interpolations for other nuclei.  All isotopes of a particular nucleus are assumed 
to have the same density.

It is well known that scattering kinematics for nucleons in a nuclear environment are 
different from those obtained in scattering from free nucleons.   
For quasi-elastic and elastic scattering, Pauli blocking is applied as described in 
Sec. \ref{sec:xsec}.  
For nuclear targets a nuclear modification factor is included to account for 
observed differences between nuclear and free nucleon structure functions 
which include shadowing, anti-shadowing, and the EMC effect \cite{Bodek:2002ps}. 

Nuclear reinteractions of produced hadrons is simulated using a cascade Monte Carlo 
which will be described in more detail in a following section. The struck nucleus
is undoubtedly left in a highly excited state and will typically de-excite by emitting
nuclear fission fragments, nucleons, and photons.  
At present de-excitation photon emission is simulated only for 
oxygen \cite{Ejiri:1993rh, Kobayashi:2005ut} due to the significance of these
3-10 MeV photons in energy reconstruction at water Cherenkov detectors.
Future versions of the generator will handle de-excitation photon emission 
from additional nuclear targets.

%
%

\subsection{Cross section model}
\label{sec:xsec}

The cross section model provides the calculation of the differential and total cross sections.  
During event generation the total cross section is used together with the flux 
to determine the energies of interacting neutrinos.  The cross sections for specific 
processes are then used to determine which interaction type occurs, and the 
differential distributions for that interaction 
model are used to determine the event kinematics. 
While the differential distributions must be calculated event-by-event, the total 
cross sections can be pre-calculated and stored for use by many jobs sharing the same
physics models.   Over this energy range neutrinos can scatter off a variety of 
different `targets' including the nucleus (via coherent scattering), individual nucleons, 
quarks within the nucleons, and atomic electrons.

     {\bf Quasi-Elastic Scattering:}  
     Quasi-elastic scattering (e.g. $\nu_\mu + n \rightarrow \mu^- + p$)  
     is modeled using an implementation of the Llewellyn-Smith 
     model \cite{LlewellynSmith:1972zm}.  In this model the hadronic weak current  
     is expressed in terms of the most general Lorentz-invariant form factors.  
     Two are set to zero as they violate G-parity.  Two vector form factors can be 
     related via CVC to electromagnetic form factors which are measured over a 
     broad range of kinematics in electron elastic scattering experiments.  
     Several different parametrizations of these electromagnetic form factors including 
     Sachs \cite{Sachs:1962zzc}, BBA2003 \cite{Budd:2003wb} and BBBA2005 \cite{Bradford:2006yz}
     models are available with BBBA2005 being the default.
     Two form factors - the pseudo-scalar and axial vector, remain.  The pseudo-scalar
     form factor is assumed to have the form suggested by the partially conserved axial current
     (PCAC) hypothesis \cite{LlewellynSmith:1972zm}, 
     which leaves the axial 
     form factor F$_A$(Q$^2$) as the sole remaining unknown quantity.  
     F$_A(0)$ is well known from 
     measurements of neutron beta decay and the Q$^2$ dependence of this form factor can
     only be determined in neutrino experiments and has been the focus of a large amount 
     of experimental work over several decades.  In GENIE a dipole form is assumed, with the 
     axial vector mass m$_A$ remaining as the sole free parameter with a default value of 0.99 GeV/c$^2$.  

     For nuclear targets, the struck  a suppression factor is included from an analytic calculation of
     the rejection factor in the Fermi Gas model, based on the simple requirement that the momentum 
     of the outgoing nucleon exceed the fermi momentum $k_F$ for the nucleus in question.  
     Typical values of $k_F$ are 0.221 GeV/c for nucleons in ${}^{12}$C, 
     0.251 GeV/c for protons in ${}^{56}$Fe, and 
     0.256 GeV/c for neutrons in ${}^{56}$Fe. 

     {\bf Elastic Neutral Current Scattering:}  
     Elastic neutral current processes are computed according to the 
     model described by Ahrens et al. \cite{Ahrens:1986xe}, where the axial form factor is given by:   
     \begin{equation}
       G_A(Q^2) = \frac{1}{2} \frac{G_A(0)}{(1+Q^2/M_A^2)^2}(1+\eta). 
     \end{equation}
     The adjustable parameter $\eta$ includes possible isoscalar contributions to the 
     axial current, and the GENIE default value is $\eta=0.12$.
     For nuclear targets the same reduction factor described above is used.

     {\bf Baryon Resonance Production:}
     The production of baryon resonances in neutral and charged current channels is 
     included with the Rein-Sehgal model \cite{Rein:1981wg}.   
     This model employs the Feynman-Kislinger-Ravndal \cite{Feynman:1971wr} model of baryon resonances, which   
     gives wavefunctions for the resonances as 
     excited states of a 3-quark system in a relativistic harmonic oscillator potential with spin-flavor symmetry.
     In the Rein-Sehgal paper the helicity amplitudes for the FKR model are computed and used to construct 
     the cross sections for neutrino-production of the baryon resonances.     
     From the 18 resonances of the original paper we include the 16 that are listed
     as unambiguous at the latest PDG baryon tables and all resonance parameters have
     been updated.    
     In our implementation of the Rein-Sehgal model interference between neighboring resonances has been 
     ignored.  Lepton mass terms are not included in the calculation of the differential cross section, but the  
     effect of the lepton mass on the phase space boundaries is taken into account.  
     For tau neutrino charged current interactions an overall correction factor to the 
     total cross section is 
     applied to account for neglected elements (pseudoscalar form factors and lepton mass terms)  in the original model.
     The default value for the resonance axial vector mass m$_A$ is 1.12 GeV/c$^2$, as determined in the global fits  
     carried out in Reference \cite{Kuzmin:2006dh}.  

     {\bf Coherent Neutrino-Nucleus Scattering:}
     Coherent scattering results in the production of forward going pions in both charged current 
     ($\nu_\mu + A \rightarrow \mu^- + \pi^+ + A$) and neutral current 
     ($\nu_\mu + A \rightarrow \nu_\mu + \pi^0 + A$) channels.  
     Coherent neutrino-nucleus interactions are modeled according to the Rein-Sehgal model \cite{Rein:1983pf}.
     Since the coherence condition requires a small momentum transfer to the target nucleus,     
     it is a low-Q$^2$ process which is related via PCAC to the pion field.  
     The Rein-Sehgal model begins from the PCAC form
     at Q$^2$=0, assumes a dipole dependence for non-zero Q$^2$, with $m_A=1.00$ GeV/c$^2$, 
     and then calculates the relevant pion-nucleus 
     cross section from measured data on total and inelastic pion scattering from protons and deuterium \cite{Yao:2006px}.   
     The GENIE implementation is using the modified PCAC formula described in a recent revision
     of the Rein-Sehgal model \cite{Rein:2006di} that includes lepton mass terms.

     {\bf Non-Resonance Inelastic Scattering:}
     Deep (and not-so-deep) inelastic scattering (DIS) is calculated in an effective leading order 
     model using the modifications 
     suggested by Bodek and Yang \cite{Bodek:2002ps} to describe scattering at low Q$^2$.  
     In this model higher twist and target mass corrections are 
     accounted for through the use of a new scaling variable and modifications to the low Q$^2$ 
     parton distributions.  
     The cross sections are computed at a fully partonic level (the ${\nu}q{\rightarrow}lq'$
     cross sections are computed for all relevant sea and valence quarks).
     The longitudinal structure function is taken into account using the Whitlow R
     ($R=F_{L}/2xF_{1}$) parameterization \cite{Whitlow:1990gk}.
     The default parameter values are those given in \cite{Bodek:2002ps}, which are determined   
     based on the GRV98 LO parton distributions \cite{Gluck:1998xa}.  
     An overall scale factor of 1.032 is applied to the predictions of the Bodek-Yang model 
     to achieve agreement with the measured value of the neutrino cross section at high energy (100 GeV).
     This factor is necessary since the Bodek-Yang model treats axial and vector form modifications 
     identically and would therefore not be expected to reproduce neutrino data perfectly.  This overall
     DIS scale factor needs to be recalculated when elements of the cross section model are changed. 

The same model can be extended to low energies; it is the model used for the nonresonant
processes that compete with resonances in the few-GeV region.

     {\bf Quasi-Elastic Charm Production:}
     QEL charm production is modeled according to the Kovalenko local duality inspired model \cite{Kovalenko:1990zi}
     tuned by the GENIE authors to recent NOMAD data \cite{Bischofberger:2005ur}.

     {\bf Deep-Inelastic Charm Production:}
     DIS charm production is modeled according to the Aivazis, Olness and Tung model \cite{Aivazis:1993kh}.
     Charm-production fractions for neutrino interactions are taken from \cite{DeLellis:2002pr}, and utilize both 
     Peterson \cite{Peterson:1982ak} 
     and Collins-Spiller \cite{Collins:1984ms} fragmentation functions, with Peterson fragmentation functions being the default.  
     The charm mass is adjustable and is set to 1.43 GeV/c$^2$ by default. 

     {\bf Inclusive Inverse Muon Decay:}
     Inclusive Inverse Muon Decay cross section is computed using an implementation of the Bardin and Dokuchaeva model \cite{Bardin:1986dk}
     taking into account all 1-loop radiative corrections. 

     {\bf Neutrino-Electron Elastic Scattering:}
     The cross sections for all ${\nu}e-$ scattering channels other than Inverse Muon Decay is computed according to \cite{Marciano:2003eq}.
     Inverse Tau decay is neglected.

%
%

\subsubsection{Modeling the transition region}
\label{sec:transition}

As discussed, for example, by Kuzmin, Lyubushkin and Naumov \cite{Kuzmin:2005bm} one typically 
considers the total ${\nu}N$ CC scattering cross section as

\begin{center}
\(\sigma^{tot} = \sigma^{QEL} \oplus
  \sigma^{1\pi} \oplus \sigma^{2\pi} \oplus ...
  \oplus \sigma^{1K} \oplus ... \oplus \sigma^{DIS} \)
\end{center}

In the absence of a model for exclusive inelastic multi-particle neutrinoproduction,
the above is usually being approximated as

\begin{center}
\(\sigma^{tot} = \sigma^{QEL} \oplus \sigma^{RES} \oplus \sigma^{DIS} \)
\end{center}

assuming that all exclusive low multiplicity inelastic reactions proceed primarily 
through resonance neutrinoproduction. 
For the sake of simplicity, small contributions to the
total cross section in the few GeV energy range, such as 
coherent and elastic ${\nu}e^{-}$ scattering, were omitted from the expression above.
In this picture, one should be careful in avoiding
double counting the low multiplicity inelastic reaction cross sections.

In GENIE release the total cross sections is constructed along the same lines, 
adopting the procedure developed in NeuGEN \cite{Gallagher:2002sf} to avoid double counting.  
The total inelastic differential cross section is computed as

\begin{center}
\(\frac{\displaystyle d^{2}\sigma^{inel}}{\displaystyle dQ^{2}dW} =
 \frac{\displaystyle d^{2}\sigma^{RES}}{\displaystyle dQ^{2}dW} +
 \frac{\displaystyle d^{2}\sigma^{DIS}}{\displaystyle dQ^{2}dW}\)
\end{center}

The term $d^{2}\sigma^{RES}/dQ^{2}dW$ represents the contribution from all
low multiplicity inelastic channels proceeding via resonance production.  This term 
is computed as

\begin{center}
\(\frac{\displaystyle d^{2}\sigma^{RES}}{\displaystyle dQ^{2}dW} =
  {\sum\limits_k} \bigl( \frac{\displaystyle d^{2}\sigma^{R/S}}{\displaystyle dQ^{2}dW} \bigr)_{k}
  \cdot {\Theta}(Wcut-W)\)
\end{center}

where the index $k$ runs over all baryon resonances taken into account,
$W_{cut}$ is a configurable parameter and $(d^{2}\sigma^{RS}_{{\nu}N}/dQ^{2}dW)_{k}$ is
the Rein-Seghal model prediction for the $k^{th}$ resonance cross section.

The DIS term of the inelastic differential cross section is expressed in terms
of the differential cross section  predicted by the Bodek-Yang model appropriately
modulated in the ``resonance-dominance" region $W<W_{cut}$ so that the RES/DIS
mixture in this region agrees with inclusive cross section data
\cite{MacFarlane:1983ax,Berge:1987zw, Ciampolillo:1979wp, Colley:1979rt, 
Bosetti:1981ip, Mukhin:1979bd, Baranov:1979sx, Barish:1978pj, Baker:1982ty, Eichten:1973cs} 
and exclusive 
1-pion \cite{Lerche:1978cp, Ammosov:1988xb, Grabosch:1988gw, Bell:1978qu, Kitagaki:1986ct,
Allen:1980ti, Allen:1985ti, Allasia:1990uy, Campbell:1973wg, Barish:1978pj,Radecky:1981fn} and 
2-pion \cite{Day:1984nf, Kitagaki:1986ct}
cross section data:

\begin{center}
\begin{eqnarray*}
   \frac{d^{2}\sigma^{DIS}}{dQ^{2}dW} &=&
   \frac{d^{2}\sigma^{DIS,BY}}{dQ^{2}dW} \cdot {\Theta}(W-Wcut) + \\
&+& \frac{d^{2}\sigma^{DIS,BY}}{dQ^{2}dW} \cdot {\Theta}(Wcut-W) \cdot {\sum\limits_m}f_{m}
\end{eqnarray*}
\end{center}

In the above expression, $m$ refers to the multiplicity of the hadronic system
and, therefore, the factor $f_{m}$ relates the total calculated DIS cross section
to the DIS contribution to this particular multiplicity channel.
These factors are computed as $f_{m} = R_{m} {\cdot} P^{had}_{m}$ where $R_{m}$ is
a tunable parameter and $P^{had}_{m}$ is the probability, taken from the
hadronization model,
that the DIS final state hadronic system multiplicity would be equal to $m$.
The approach described above couples the GENIE cross section and hadronic multiparticle
production model \cite{Yang:2007zzt}.   

%
%

\subsection{Neutrino-induced hadronic multiparticle production modeling}

Neutrino-induced hadronic shower modeling is an important aspect of the
intermediate energy neutrino experiment simulations, as non-resonant inelastic
scattering becomes the dominant interaction channel for neutrino energies
as low as 1.5 GeV.

Experiments are sensitive to the details of hadronic system modeling in many 
different ways.  
Experiments making calorimetric measurements of neutrino energy in charged
current reactions are typically calibrated using single particle test beams, which introduces a model dependence in
determining the conversion between detector activity and the energy of neutrino-produced hadronic systems.
Physics analysis can also depend on the prediction of the hadron shower characteristics,
such as shower shapes, energy profile and particle content, primarily for event identification.
A characteristic example is a $\nu_{\mu} \rightarrow \nu_{e}$ appearance analysis, where
the evaluation of backgrounds coming from neutral current (NC) events, would be quite sensitive on the details
of the NC shower simulation and specifically the $\pi^{0}$ shower content.
It is therefore imperative that the state-of-the-art in shower modeling is included in our
neutrino interaction simulations.

GENIE uses the AGKY hadronization model which was developed for the MINOS experiment \cite{Yang:2007zzt}. 
This model integrates an empirical low-invariant mass model with PYTHIA-6 at higher
invariant masses.  The transition between these two models takes place over an adjustable 
window with the default values of 2.3 GeV/c$^2$ to 3.0 GeV/c$^2$, so as to ensure continuity of all simulated
observables as a function of the invariant mass.  For the hadronization of low-mass states the 
model proceeds in two phases, first 
determining the particle content of the hadronic shower, and  secondly  determining the 4-momenta of 
the produced particles in the hadronic center of mass. 

The AGKY's low mass hadronization model generates hadronic systems that typically
consist of exactly one baryon ($p$ or $n$) and any number of $\pi^{+}$, $\pi^{-}$, $\pi^{0}$,
$K^{+}$, $K^{-}$, $K^{0}$, ${\bar{K^{0}}}$ mesons kinematically possible and allowed by 
charge conservation.

For a fixed hadronic invariant mass and initial state (neutrino and hit nucleon), the
algorithm for generating the hadron shower particles generally proceeds as follows:
\begin{itemize}
\item Compute the average charged hadron multiplicity $<n_{ch}>$ for a given invariant mass (W) 
     using empirical expressions of the
     ($<n_{ch}>=a_{ch}+b_{ch}*lnW^{2}$) form. 
     The coefficients $a_{ch}$, $b_{ch}$, which depend on the initial state,
     have been determined by bubble chamber experiments \cite{Zieminska:1983bs}. 
\item Compute the average total hadron multiplicity $<n_{tot}>$ as $<n_{tot}>=1.5<n_{ch}>$.
\item Using the calculated average hadron multiplicity, generate the actual
     hadron multiplicity taking into account that the multiplicity dispersion is
     described by the KNO scaling law, ($<n>P(n)=f(n/<n>)$) \cite{Koba:1972ng}. 
     P(n) is the probability of having $n$ hadrons in the  
     final state given an expected average of $<n>$, and $f(n/<n>)$ is a universal scaling function. 
     The KNO scaling is parametrized employing the Levy
     \footnote{The Levy function $Levy(z;c) = 2e^{-c}c^{cz+1}/\Gamma(cz+1)$}
     function with an input parameter $c_{ch}$  that depends on the initial state and is
     treated as a tuning parameter.
\item Generate hadrons up to the generated hadron multiplicity taking into account the hadron
     shower charge conservation and the kinematical constraints.
     Protons and neutrons are produced in the ratio
     2:1 for $\nu p$ interactions, 1:1 for $\nu n$ and $\bar{\nu}p$, and 1:2 for $\bar{\nu}n$ interactions.
     Charged mesons are then created in order to balance charge, and the remaining  mesons are generated in neutral pairs. 
     The probabilities for each are 31.33\% ($\pi^{0},\pi^{0}$), 62.67\% ($\pi^{+},\pi^{-}$)
     and 6\% of strange meson pairs.
     The probability of producing a strange baryon via associated production is determined from a 
     fit to $\Lambda$ production data 
     \cite{Jones:1992bm,Baker:1986xx,DeProspo:1994ac,Bosetti:1982vk}:
     \begin{equation}
       P_{hyperon}= a_{hyperon}+ b_{hyperon}\  \ln W^{2}
     \end{equation}
\end{itemize}

\begin{figure}[ht]
\begin{minipage}[b]{0.47\linewidth}
\centering
\includegraphics[width=17pc]{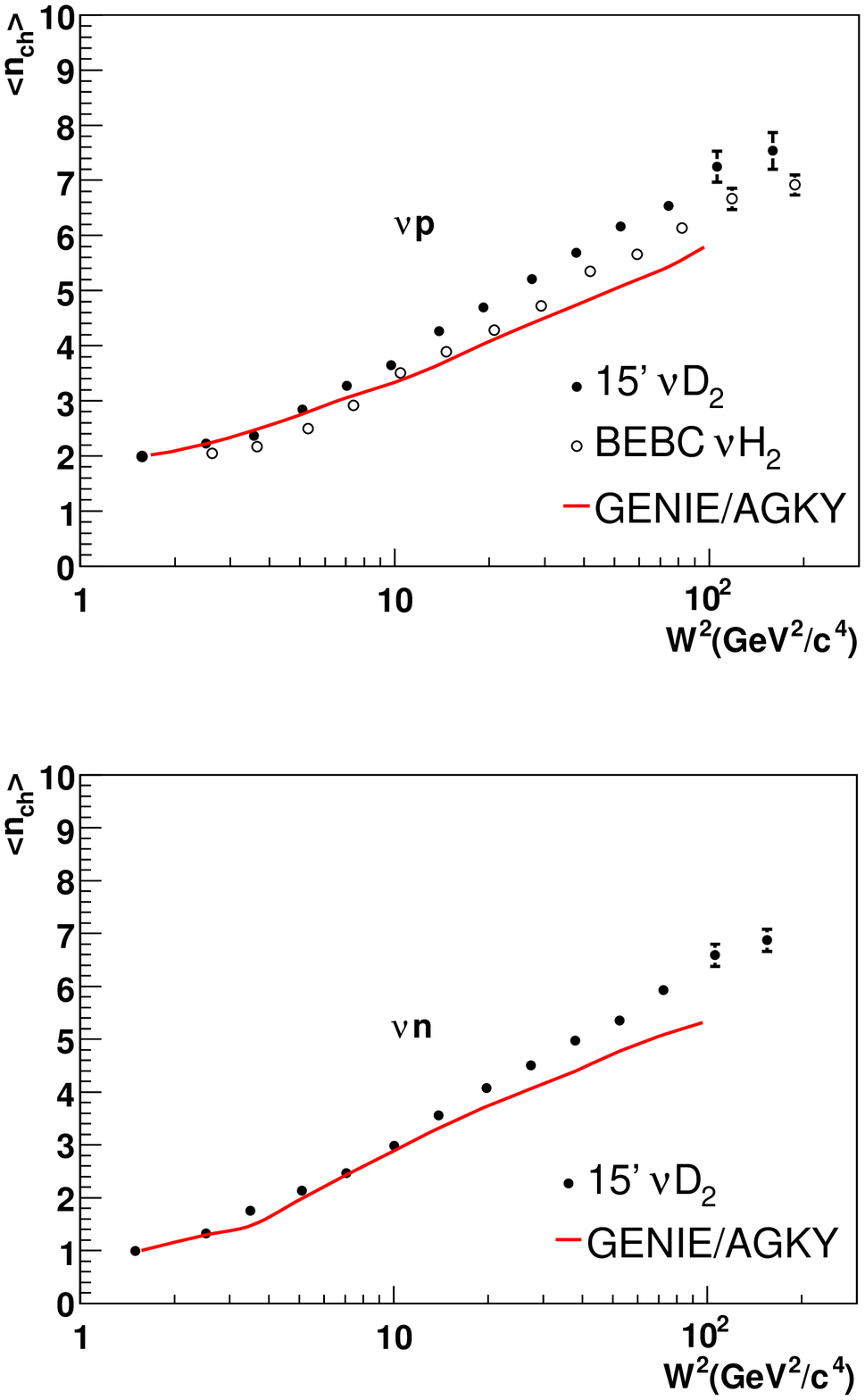}
\caption{
 Data/model comparisons of the average charged hadron multiplicity $<n_{ch}>$
 shown as a function of the squared hadronic invariant mass for
 $\nu p$ and $\nu n$ interactions. 
 Data are from Refs. \cite{Zieminska:1983bs, Allen:1981vh}.
}
\label{fig:ChMult}
\end{minipage}
\hspace{0.5cm}
\begin{minipage}[b]{0.47\linewidth}
\centering
\includegraphics[width=17pc]{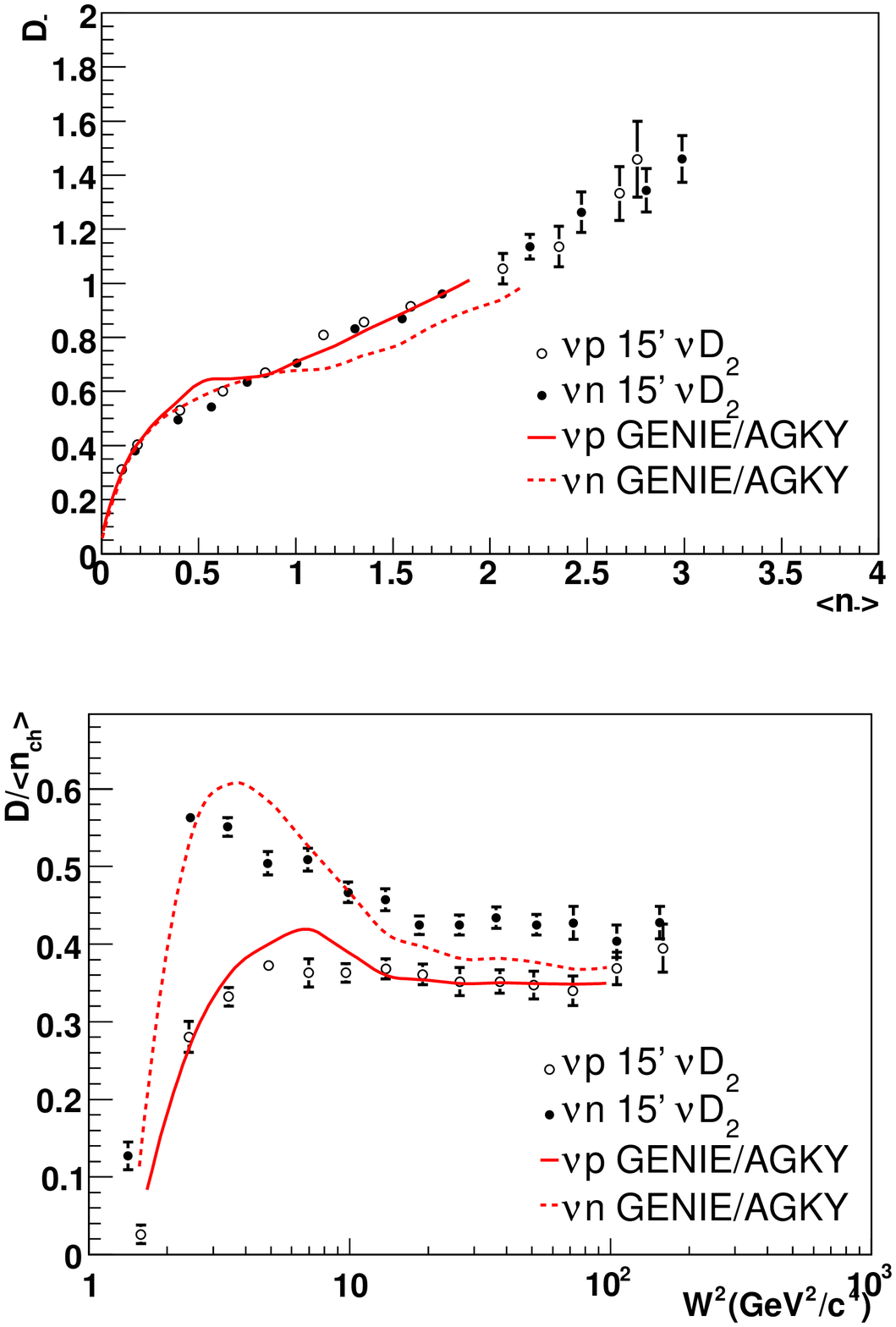}
\caption{
 Data/model comparisons of the negatively charged hadron multiplicity
 dispersion $D_{-}$ as a function of the average charged hadron multiplicity $<n_{-}>$
 (up) and the reduced dispersion $D_{-}/<n_{-}>$ as a function of the squared hadronic
 invariant mass (down).
 Data are from Ref. \cite{Zieminska:1983bs}.
}
\label{fig:ChD}
\end{minipage}
\end{figure}

Fig. \ref{fig:ChMult} shows the 
data/model comparisons of the average charged hadron multiplicity $<n_{ch}>$
as a function of the squared hadronic invariant mass for $\nu p$ and $\nu n$ interactions.
Fig. \ref{fig:ChD} shows the
data/model comparisons of the negatively charged hadron multiplicity dispersion $D_{-}$ 
as a function of the average charged hadron multiplicity $<n_{-}>$
and the reduced dispersion $D_{-}/<n_{-}>$ as a function of the squared hadronic
invariant mass.

The main dynamical feature observed in the study of hadronic systems is that the baryon
tends to go backwards in the hadronic center of mass and 
that the produced hadrons have small transverse momentum relative to the direction of momentum transfer.   
These features are naturally described in the quark model where the baryon is formed from the diquark 
remnant, which goes backwards in the center-of-mass, and transverse momentum is generated solely 
through intrinsic parton motion and gluon radiation.    At low invariant masses energy-momentum constraints on the
available phase space play a particularly important role.  
The most pronounced kinematical feature in this region is that one of the
produced particles (proton or neutron) is much heavier than the rest (pion and kaons) and
exhibits a strong directional anticorrelation with the momentum transfer.

Our strategy, therefore, is to correctly reproduce the final state nucleon momentum,
and then perform a phase space decay on the remnant system employing, in addition, 
a $p_T$-based rejection scheme designed to reproduce the expected meson transverse momentum distribution. 
The nucleon momentum is generated using input $p_{T}^{2}$ and $x_{F}=(p_L^*/p_{L max}^*)$ PDFs which are parametrized
based on experimental data \cite{Derrick:1977zi,CooperSarkar:1982ay}.  Once the baryon momentum is selected the remaining particles 
undergo a phase space decay.
The phase space decay employs a rejection method suggested in \cite{Clegg:1981js},
with a rejection factor $e^{-A*p_{T}}$ for each meson. This causes the transverse momentum distribution of the generated mesons to fall exponentially with increasing $p^{2}_{T}$, controlled by the adjustable parameter $A$ which has a default value of 3.5 GeV$^{-1}$.   
Here $p_{T}$ is the momentum component perpendicular to the current direction.
One of the remaining challenges in this model, which will be addressed in the future, is to better describe forward and 
backward hemisphere multiplicity distributions.  The forward/backward multiplicity distributions yield an unphysically rapid 
transition to the PYTHIA values, a feature not seen in other recent hadronization models \cite{Nowak:2006xv}.

\begin{figure*}[htb]
\includegraphics[width=35pc]{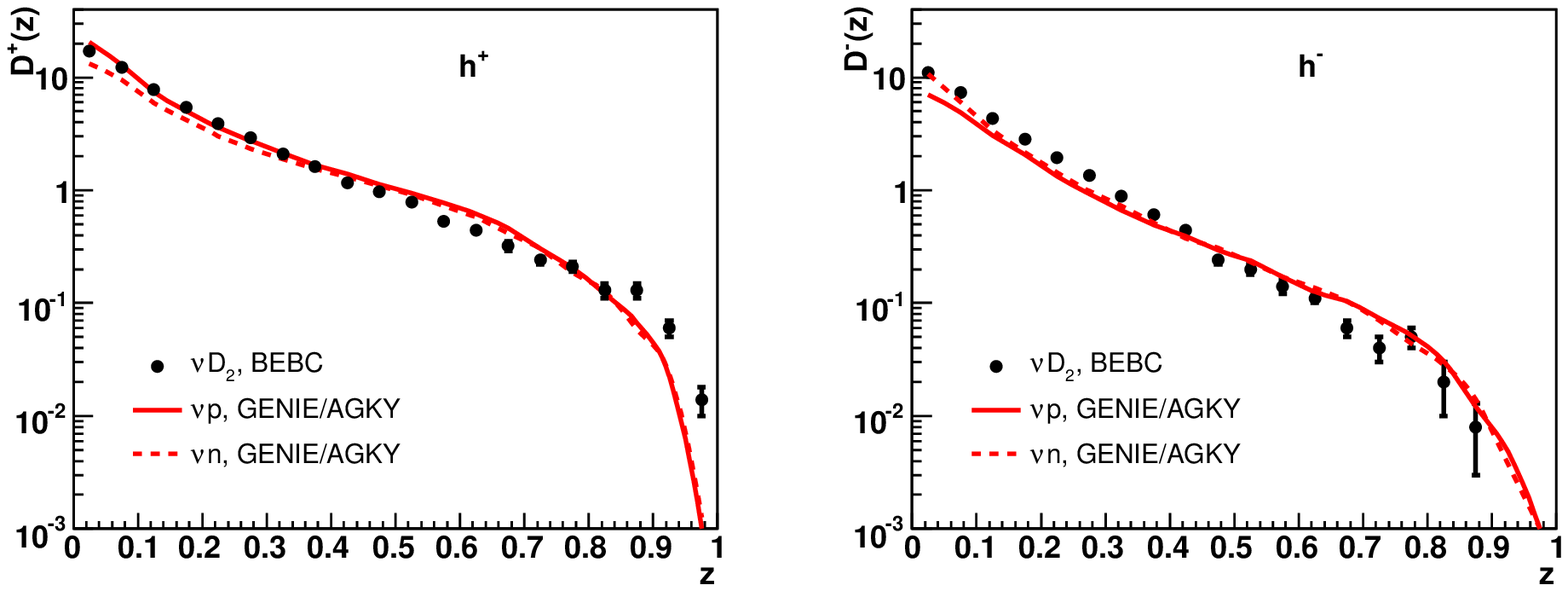}
\caption{
Data/model comparisons of the fragmentation function for 
positively and negatively charged hadrons. 
Applied cuts: Squared hadronic invariant mass $W^{2}$ above 5 $GeV^{2}/c^{4}$
and squared 4-momentum transfer $Q^{2}$ above 1 $GeV^{2}/c^{2}$.
Data are from Ref. \cite{Allasia:1984ua}.
}
\label{fig:ZFragm}
\end{figure*}

Fig. \ref{fig:ZFragm} shows the data/model comparisons of the fragmentation function for positively and negatively
charged hadrons.
2-body hadronic systems are a special case:
The hadronic system 4-momenta are generated by a simple unweighted phase space.

%
%

\subsubsection{Intranuclear rescattering}

The hadronization model describes particle production from free targets and is 
tuned primarily to bubble chamber data on hydrogen and deuterium 
targets \cite{Zieminska:1983bs, Derrick:1977zi, Allen:1981vh, 
Ivanilov:1984gh, Grassler:1983ks, Allasia:1984ua, Berge:1978fr, Ammosov:1978vt}.
Hadrons produced in the nuclear environment may rescatter on their way out of the 
nucleus, and these reinteractions significantly modify the observable distributions
in most detectors.  

It is also well established that hadrons
produced in the nuclear environment do not immediately reinteract with their
full cross section.  The basic picture 
is that during the time it takes for quarks to materialize as 
hadrons, they propagate through the nucleus with a dramatically 
reduced interaction probability.  
This was implemented in 
GENIE as a simple `free step' at the start of the 
intranuclear cascade during which no interactions 
can occur.  The `free step' comes from a formation time of 0.523
fm/c according to the SKAT model \cite{Baranov:1984rv}. 

Intranuclear hadron transport in GENIE is handled by a subpackage called INTRANUKE. 
INTRANUKE is an intranuclear cascade simulation and has gone through numerous 
revisions \cite{dytman-ladek} since the original version was developed for use by the Soudan 2 
Collaboration \cite{Merenyi:1992gf}. The sensitivity of a particular experiment
to intranuclear rescattering depends strongly on the detector technology, 
the energy range of the neutrinos, and the physics measurement being made. 
INTRANUKE simulates rescattering of pions and nucleons in the nucleus.  

In principle one would like to have a fully realistic nuclear model which 
accurately describes the full range of processes to model particle production with
energies as low as 1 MeV to ensure
that the simulations are suitable for any conceivable experiment. Nuclear
simulations of this type are themselves highly complex packages and the integration
of these packages with GENIE is an area of active work.  An alternative approach
is to develop a simpler nuclear model, in the context of a particular experiment, 
and ensure that the relevant physics for that experiment are correctly described. 
This approach has the advantage of yielding a far simpler code, which is understood
by the experimenters.  This has particular advantages for the study of systematic
errors and the development of ancillary code like reweighting packages. 

The current version was optimized for use by the 
MINOS experiment.  For this experiment the task of developing an intranuclear
rescattering model is simplified because the detector is composed largely
of a single element, iron,
and the detector is designed to make a calorimetric
energy measurement rather than track individual particles.  
For the oscillation measurement of MINOS \cite{Michael:2006rx} the primary goal is 
ensuring that the `missing energy' lost in the nuclear environment is being reliably 
simulated.  The model has applicability to almost all nuclei and a
wide range of energies.

To handle a wide energy range for neutrinos, GENIE has defined processes
for hadrons up to 1.8 GeV kinetic energy in terms of all the relevant cross sections.  
For higher hadron energy, the underlying cross sections are assumed to be constant
at the 1.8 GeV value.  This is a good approximation to the actual values.
The code then has a description of all hadrons coming from neutrinos at all relevant energies.

\begin{figure*}[htb]
\includegraphics[width=17pc]{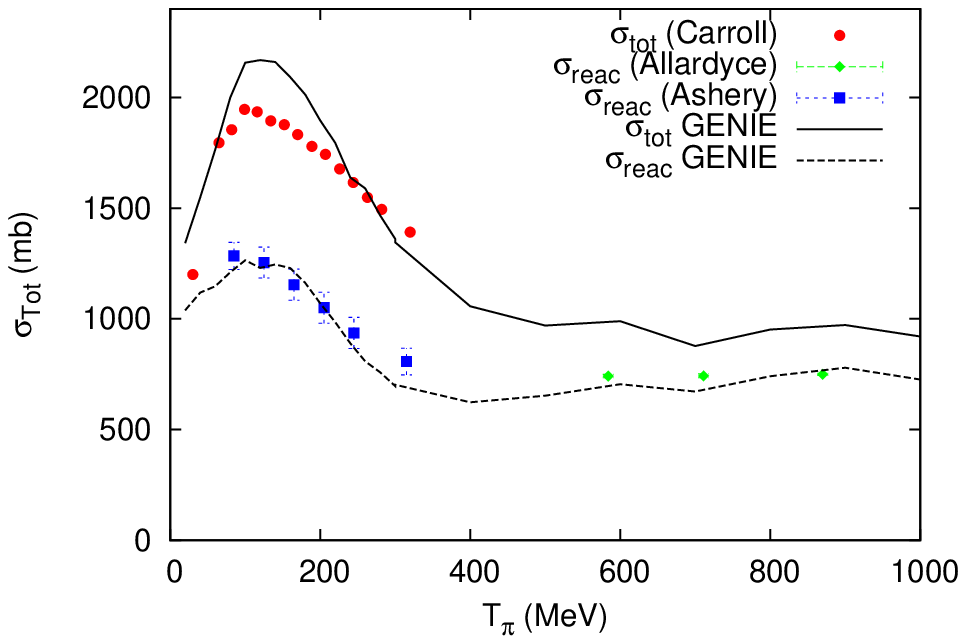}
\includegraphics[width=17pc]{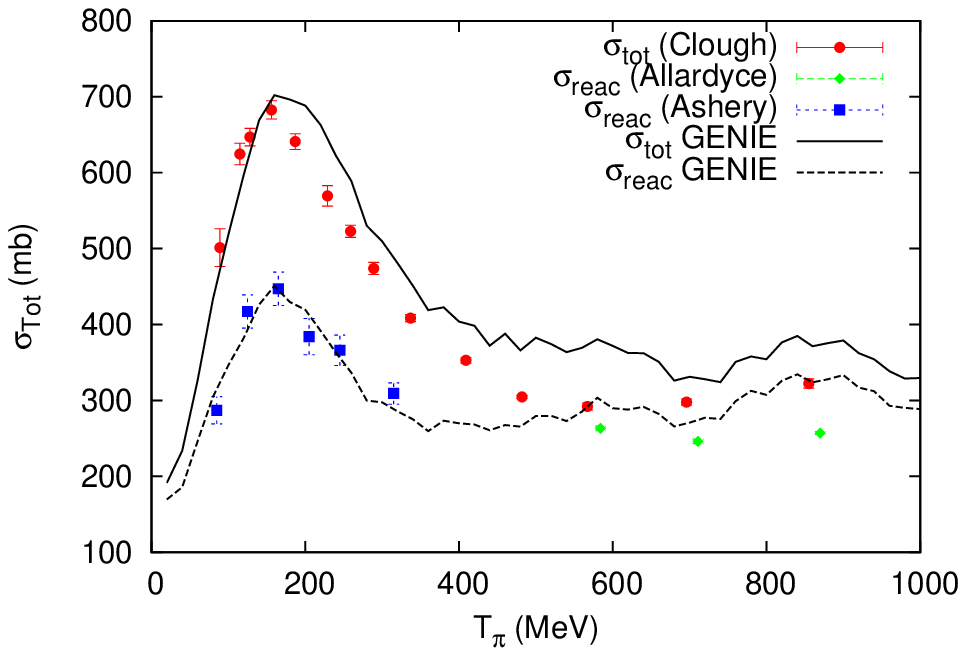}
\caption{$\pi^+$ total and reaction cross sections for iron (left) and
carbon (right). Data are from Refs. \cite{Ashery:1981tq,Allardyce:1973ce,Carroll:1976hj,Clough:1974qt}. }
\label{fig:pife}
\end{figure*}

The simulation tracks particles through the nucleus in steps of 0.05 fm.  For each 
particle only one reinteraction is allowed, and the simulation consists of the 
following steps: 
\begin{enumerate}
\item
{\bf Mean Free Path:}  In order to determine if the particle interacts in a 
particular step the mean free path ($\lambda$) is calculated based on the local density
of nucleons ($\rho(r)$) \cite{DeJager:1987qc} and a partial wave analysis of the large body of 
hadron-nucleon cross sections ($\sigma_{hN}$) \cite{Arndt:2003if}:
\begin{equation}
\lambda(r,E_h)=\frac{1}{\sigma_{hN}(E_h)\rho(r)}
\end{equation}.  
Here, $\sigma_{hN}$ is the isospin averaged cross section for the
propagating hadron (with energy $E_h$) interacting with a nucleon and $\rho(r)$ is the
matter density of nucleons at the position of the propagating hadron.
We use charge densities which are well-measured and are known to be very similar
to the matter densities.  The code presently tracks pions and nucleons.
Isospin relations are used to estimate $\pi^0-$nucleon reactions.
All nuclei heavier than oxygen are modeled with a Woods-Saxon
density distribution and lighter nuclei are modeled with a
modified Gaussian distribution.  

One difficulty in this approach is that our treatment is using a semiclassical 
model to describe a quantum mechanical process.  This poses particular difficulty 
in describing elastic scattering which dominates the total cross section at 
low energy.  This wave/particle distinction depends on energy 
with lower energy hadron-nucleus scattering being more wave-like. To account for this
we increase the size of the nuclear density distribution through which the 
particle is tracked by an amount 
\begin{equation}f\frac{hc}{p},\end{equation}
where $f$ is an adjustable dimensionless parameter set to 0.5 for pions and
1.1 for nucleons in the current default.  
We use the isospin-averaged total cross sections for pions and nucleons. 
\item
{\bf Determining the Interaction Type:}
Hadron-nucleus interactions occur with different processes and each has an associated
cross section - $\sigma_{elas}$ for elastic scattering (residual nucleus in the ground state),
$\sigma_{inel}$ for inelastic
scattering (residual nucleus in excited state, typical response is single nucleon emission),
$\sigma_{cex}$ for single charge exchange (outgoing hadron changes charge, typical response
is single nucleon emission) for all hadrons.  For pions, emission of 2 or more nucleons with no pion in
the final state is called absorption - $\sigma_{abs}$; for nucleons, a final state
with 3 or more nucleons is called multi-nucleon knockout - $\sigma_{ko}$.
For low energy nucleons, the knockout mechanism dominates.
At higher energies (above about 400 MeV for pions and 600 MeV for nucleons),
the probability of pion production ($\sigma_{piprod}$) becomes important.
The total cross section ($\sigma_{tot}$) is the sum of all component cross sections
and the reaction cross section ($\sigma_{reac}$) is the sum of cross section for
all inelastic reactions,
\begin{equation}
\sigma_{reac}=\sigma_{cex}+\sigma_{inel}+\sigma_{abs}+\sigma_{piprod}=\sigma_{tot}-\sigma_{elas}.
\end{equation}
This equation is specific to pions; $\sigma_{abs}$ is replaced by $\sigma_{ko}$ for nucleons.
Once it has been determined that a hadron reinteracts in the nucleus, the type of the
interaction is determined based on the measured cross sections for the above listed
processes.  Cross sections for kinetic energy less than 1 GeV are used and they
are assumed constant above 1 GeV.  Where data is sparse, cross section estimates are taken
from calculations of the CEM03 group \cite{Mashnik:2005ay}.  Since only 1 reinteraction
is allowed, the effect of additional reactions with the rest of the nucleus must be
included here.  The distribution of final states is optimized for iron, but valid
for all targets.  Figure~\ref{fig:pife} shows the default INTRANUKE
compared with $\pi^+$ data for $\sigma_{tot}$ and $\sigma_{reac}$ for iron and carbon.  
These are examples of many successful comparisons.

\item
{\bf Final State Products:}
Once the interaction type has been determined, the four-vectors of final 
state particles need to be generated.  Where possible these distributions 
are parametrized from data or from the output of more sophisticated nuclear 
models \cite{Mashnik:2002mw}.  Simple models are used for elastic scattering angular 
distributions.  The quasielastic reaction mechanism is known
to dominate the final state for inelastic or charge exchange processes.
Fermi motion and binding energy are used to get a good description
of the kinematics.  Whenever 3 or more particles are emitted,
distributions are according to phase space.
For MINOS, the most important issue is missing energy generated by 
inelastic and absorption processes.  Very low energy
hadrons and nuclear recoils are not seen, so simplifications can be made.  
All states where more than 5 nucleons are emitted are treated
as though 5 nucleons (3 protons and 2 neutrons) were emitted.
These restrictions are relaxed in the most recent code versions to
better match what is seen in data.

\end{enumerate}

\subsection{Physics Model Tuning}
\label{sec:tuning}

The full range of models involve more than one hundred adjustable parameters, the complete set of 
which are given in the Physics and User Manual \cite{genie-doc-753}.  
Only the most important in the construction 
of the physics models will be discussed here.  
Electroweak parameters and
CKM matrix elements use the values of the Particle Data Group \cite{Yao:2006px}.  

\begin{figure}[htb]
\center
\includegraphics[width=30pc]{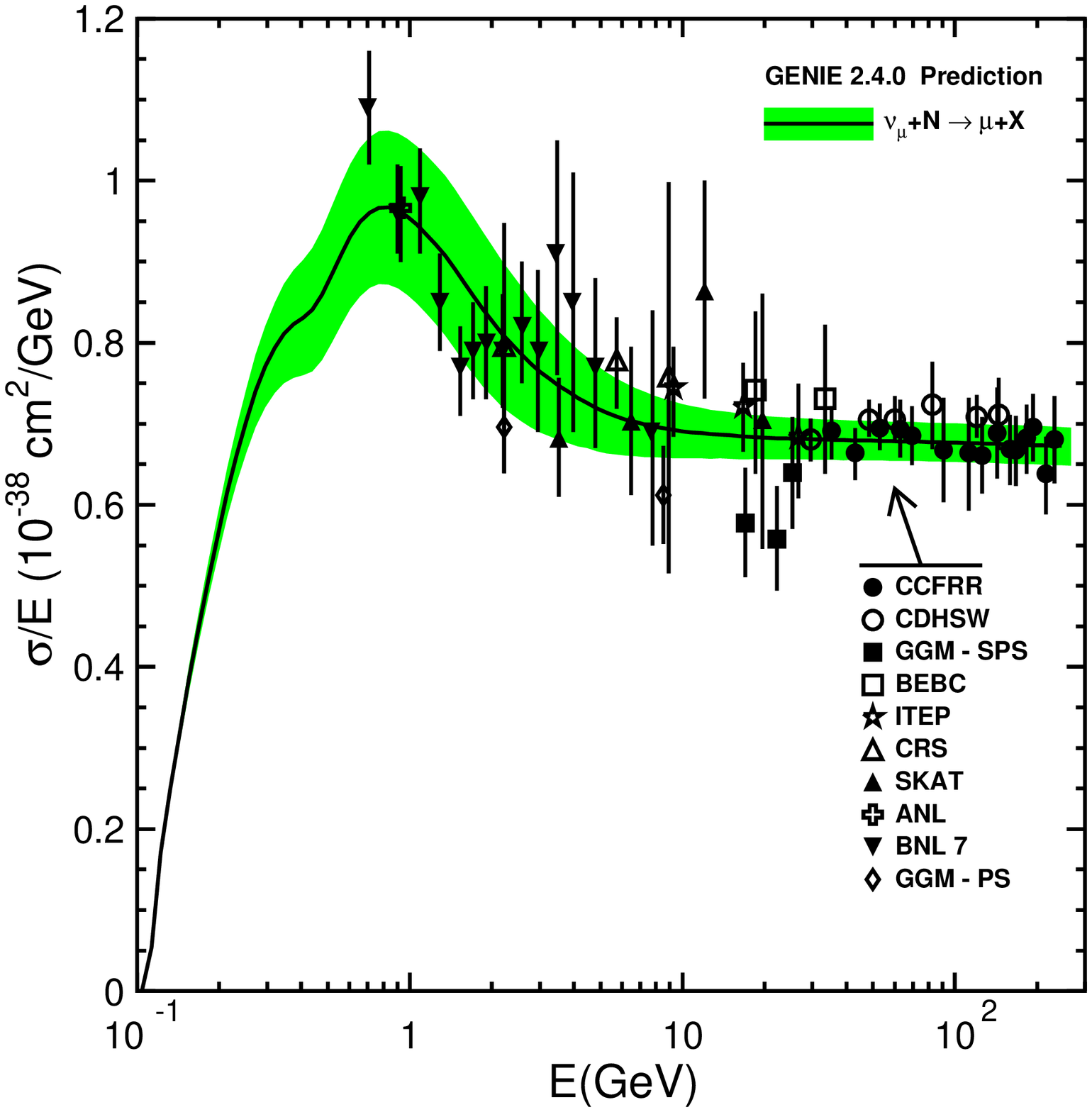}
\caption{GENIE default cross section for $\nu{\mu}$ charged current scattering 
from an isoscalar target.  The shaded band indicates the estimated 
uncertainty on the free nucleon cross section.
Data are from 
\cite{MacFarlane:1983ax} (CCFRR), 
\cite{Berge:1987zw} (CDHSW), 
\cite{Ciampolillo:1979wp} (GGM-SPS),
\cite{Colley:1979rt, Bosetti:1981ip} (BEBC),
\cite{Mukhin:1979bd} (ITEP),
\cite{Baranov:1979sx} (CRS, SKAT),
\cite{Barish:1978pj} (ANL),
\cite{Baker:1982ty} (BNL) and
\cite{Eichten:1973cs} (GGM-PS)
}
\label{fig:XSecErrEnvelope}
\end{figure}

As mentioned previously, the quasi-elastic, resonance production, and DIS models employ 
form factors, axial vector masses, and other parameters which have been determined by 
others in their global fits \cite{Bradford:2006yz,Kuzmin:2006dh}.  In order to check
the overall consistency of our model, and to verify that we have correctly implemented
the DIS model, predictions are compared to electron scattering inclusive data \cite{Gallagher:2004nq,e49a10}
and neutrino structure function data \cite{Bhattacharya:2009zz}.  
The current default values for transition 
region parameters are  $W_{cut}$=1.7 GeV/c$^2$, $R_2(\nu p)=R_2(\overline{\nu}n)$=0.1,
$R_2(\nu n)=R_2(\overline{\nu}p)$=0.3,
and $R_m$=1.0 for all $m>2$ reactions.  These are determined from fits to inclusive and exclusive 
(one and two-pion) production neutrino interaction channels.  For these comparisons we rely heavily
on online compilations of neutrino data \cite{Whalley:2004sz} and related fitting tools \cite{Andreopoulos:2005vd}
that allow one to include some correlated systematic errors (such as arising from flux uncertainties).  
The GENIE default cross section for $\nu{\mu}$ charged current scattering
from an isoscalar target,
together with the estimated uncertainty on the total cross section, as evaluated in \cite{Adamson:2007gu}  are shown 
in Fig. \ref{fig:XSecErrEnvelope}.  

The tuning of the hadronization model is accomplished using data from the BEBC \cite{Allen:1981vh}, FNAL \cite{Derrick:1981br}, 
and SKAT \cite{Baranov:1983st} bubble chamber experiments, and is described in more detail elsewhere \cite{Yang:2009zx}. 
Multiplicity measurements include
averaged charged and neutral particle ($\pi^{0}$) multiplicities, 
forward and backward hemisphere average multiplicities and correlations,topological cross sections of charged particles, and neutral - charged pion multiplicity correlations.Hadronic system measurements include fragmentation functions ($z$ distributions),
       $x_{F}$ distributions,
       $p^{2}_{T}$ (transverse momentum squared) distributions, and 
       $x_{F} - \langle p_{T}^{2} \rangle$ correlations (``seagull'' plots) \cite{Schmitz:1981sg}.    
Averaged charged particle multiplicity and 
dispersion parameters are taken from published values \cite{Zieminska:1983bs}, 
as well as our own fits \cite{Yang:2009zx}.  Baryon 4-momentum distributions 
are determined from fits to experimental data \cite{Derrick:1977zi,CooperSarkar:1982ay}.   
The settings for PYTHIA parameters are taken as the non-default
values tuned for the NUX \cite{Rubbia:2001} generator, a high energy generator used by 
the NOMAD \cite{Astier:2003gs} experiment.  


The intranuclear rescattering model has been tested and tuned based on 
comparisons to hadron-nucleus data. Hadron-nucleus cross sections are
calculated by `MC experiments' where a nucleus is being illuminated by 
a uniform hadron beam with transverse radius larger than the nucleus size.
Figure \ref{fig:pife} shows the 
comparison between INTRANUKE and data for $\pi^+-$C and $\pi^+-$Fe total and reaction 
cross sections.  Extrapolations to higher energy are required in many cases
as data for only $\sigma_{reac}$ is available.  CEM03 \cite{Mashnik:2002mw}
results with appropriate rescaling to match data at lower energies are often used.
Although the model is tuned to hadron scattering on iron, the simplicity 
of the Fermi Gas model and the  A$^{(2/3)}$ scaling of the cross sections
allow the model to be applied to nearly all nuclei encountered in the simulation 
as well. 

Validation of the intranuclear rescattering model 
using neutrino data has also been performed.  This revisits the analysis 
of Reference \cite{Merenyi:1992gf} which compares ANL neutrino scattering data on deuterium \cite{Radecky:1981fn}
to BEBC neutrino-neon data \cite{Angelini:1986gx}, where each have been rescaled to an atmospheric
neutrino flux.  By comparing neon and deuterium final state topology fractions the 
rates for pion absorption and charge exchange in neutrino-neon interactions can be determined.   
INTRANUKE reproduces the measured final state topology fractions  with an overall $\chi^2$ 
of 16.0 for 12 degrees of freedom.  The rates of pion absorption and charge exchange produced by 
INTRANUKE in this comparison are $18.3\pm 0.5$\% and $2.9\pm0.2$\% respectively, in good agreement with the measured
values of $22\pm5$\% and $10\pm8$\%.

In evaluating the uncertainty in the intranuclear rescattering model, several sources of 
uncertainty were taken into account for their effect on the MINOS determination of the 
hadronic energy scale \cite{Dytman:2008st}.  These include the experimental uncertainty on the
external data that serves as the input to the model, as well as on some of the key theoretical 
assumptions in the model, in particular the modeling of pion absorption reactions and the 
treatment of low energy pion scattering \cite{Dytman:2008st}.

\section{Recent Developments}
\label{future}
Recent focus on development for GENIE has been on the nuclear structure
and final state interaction codes.  The purpose is to make the code
well-tuned to the needs of 
the upcoming generation of accelerator experiments including T2K \cite{Wark:2008zz},
NoVA \cite{Ayres:2004js}, Minerva \cite{Schulte:2006db},
MicroBooNE \cite{Soderberg:2009rz} and ArgoNEUT \cite{Soderberg:2009qt}.
We are also looking to push the validity range of GENIE down to the MeV scale
making it applicable for the study of neutrinos from reactors, supernova and
SNS \cite{Scholberg:2007zz}.

The Fermi Gas nuclear model has been shown to be wrong in detail through
interpretation of electron scattering experiments \cite{Frois:1987hk}.
Nucleon-nucleon correlations are important in kinematic regions where
the impulse approximation is unlikely to apply.  The spectral function
\cite{Benhar:2006wy,Ankowski:2007uy} has become a useful model to represent the
effects of a many-body model.  Developmental versions of GENIE now
contain spectral functions of Benhar for carbon, iron, and lead.
Code for calculating $(e,e^\prime)$ differential cross sections is
now in place.

A true internuclear cascade model has also been in development \cite{dytman-ladek}.
It tracks pions and nucleons through multiple reactions in the
same nucleus in which the neutrino was absorbed.  Free hadron-nucleon 
cross sections are used with the struck nucleon has momentum and binding
energy according to the Fermi Gas model.  Interactions for protons, neutrons, and pions
are presently modeled.  The hadrons are
on-shell between scatterings; the reactions are governed by
the same mean free path and Fermi gas models as for the existing
model.  Thus, there is no limit on the number of particles that
can be tracked.  A simple model for compound nuclear processes
is included to properly account for effects in hadron-nucleus data.

\section{Software Design Overview} 
\label{sec:design}

In this section we will describe the software design of the GENIE package.   We begin by discussing  
the software requirements and use cases. The software framework is presented together with key classes 
such as \em particles, events, \em and \em interactions\em. The hierarchical delegation of responsibility during event 
generation to \em driver, thread, \em and \em module \em classes is described. 

\subsection {Requirements}

The process of requirements capture and software design was carried out over a period 
of several months in 2004 and 2005. 
During this phase there was extensive discussion within the MINOS collaboration as well as in 
conjunction with the NuINT\footnote{Neutrino Interactions in the Few-GeV Energy Range} Conference series.    
Through these meetings we received input from many users of these packages as well as with several 
experts who had designed, developed, and supported previous (Fortran-based) procedural codes.
   
Through these discussions the need for a package of this type for future neutrino experiments became very clear.    
The procedural programs had often been in use for several decades and had expanded greatly beyond their initial scope.   
They had often reached a critical mass where further modifications were deemed extremely difficult because of the 
overall fragility of the architecture.  In addition there were often strong couplings between aspects of the simulation 
that made incremental improvements in a particular area difficult.  Another commonly voiced concern was the lack 
of documentation, particularly regarding the ways in which the models and their parameters were tuned to data.   

These discussions served to illuminate several typical use cases for neutrino event generators and related tools:

\begin{itemize}
\item
For event generation in conjunction with a full detector simulation. 
\item
For event generation in fast simulations, either 4-vector only or using a parametrized model of the detector response.    
\item
To provide a library of cross section values for interaction rate calculations. 
\item
As a source of information about the underlying models and their uncertainties. 
\item
As a primary tool in the evaluation of systematic uncertainties.  
\end{itemize}

These use case evaluations and discussions led to the establishment of a set of requirements 
for the overall architecture as well as requirements for the documentation and ancillary tools:

\begin{itemize}
\item
Decouple physics models as much as possible from the framework code.  
\item
Lower the barrier to entry for physics model developers (particularly theorists).
\item
Provide a well-tested, well-defined default configuration which provides a benchmark 
for all users who are primarily interested in using the package in black box-mode. 
\item
Incorporate up-to-date theoretical and experimental work and provide a flexible 
framework so that it can be maintained. 
\item
Incorporate state-of-the-art software engineering methodologies to support 
these goals through an object-oriented design. 
\item
Leverage the developments in other areas of HEP software development 
(in particular ROOT class libraries).  
\item
Provide the external data used to tune the package as part of the overall distribution.  
\item
Provide clear documentation about how the models are tested, tuned, and validated.  
\item
Provide a set of tools to facilitate the tuning of model parameters and for an independent 
evaluation of how well models describe existing data.  
\end{itemize}

These requirements were met in the August 2007 first release of the GENIE package.
The implementation of the physics models was cross-checked through an exhaustive 
set of comparisons \cite{genie-doc-755} with one of the existing procedural codes \cite{Gallagher:2002sf}.   
GENIE can be configured to be identical in its physics 
content to the final version (v3.5.5) of the NEUGEN3 package, one of the legacy procedural codes
which had been used by numerous experiments for over more than a decade.


\subsection{Core Framework}
\label{sec:core}

The key requirement of the GENIE Core Framework is to transparently decouple 
the high-level code focusing on physics simulations from the low-level structures 
involved primarily with memory management and configuration.

The framework was developed and reviewed primarily within the MINOS experiment and, 
inevitably, has been influenced by the MINOS offline software design \cite{MINOSOffline:PrivCom}.
In developing the GENIE framework we recycled, adapted and extended key 
features of the MINOS offline framework.  We drew heavily from the accumulated software 
engineering experience encapsulated within popular software design patterns,
including the Visitor, Chain of Responsibility, Factory, Strategy and Singleton \cite{DesignPatterns:GangOfFour}.
The Core Framework is not specific to the subject matter domain of GENIE and could 
be adapted and reused in other scientific computing applications.
The GENIE Event Generation Framework, to be discussed later, 
is a subject matter-specific layer built on top of the Core Framework.

The framework concerns itself mainly with the properties, instantiation and memory 
management of software abstractions called \em Algorithms\em, 
and specifies the interfaces that underpin the interactions between the numerous 
concrete \em Algorithm \em realizations.
The \em Algorithm \em is a key Core Framework abstraction. The notion of an `algorithm' 
in an object-oriented system requires further clarification as it does not correspond to
its more familiar notion in the context of procedural software systems.
In the Core Framework the \em Algorithm \em encapsulates the common
behavior of all algorithmic objects. It is an abstract base class which defines exactly 
how algorithmic objects are to be initialized and configured,
how they are to look up their configuration, 
how they are to be identified, and how they report their status. 
These are common, largely operational features that characterize a very heterogeneous 
collection of algorithms such as  
cross section models, hadronization models, particle decayers, form factor and
structure function models, event generation modules and threads and other types
of algorithms that can be found within GENIE.
The fact that such a common behavior is imposed upon all algorithmic objects 
allows us to build a central, external, XML \cite{XML:Spec} - based algorithm configuration 
system that contributes significantly to the flexibility and extensibility of GENIE.
The kind of computation to be performed, the usual identifying feature of an algorithm 
in a procedural system, is a secondary characteristic at this level of abstraction.

At the next level up from the \em Algorithm \em root of an algorithm inheritance tree, 
we find a standardized interface which defines how to invoke each specialized type of
calculation and retrieve its results.
Numerous such specialized algorithm interfaces exist within GENIE.
Examples include the \em GFluxI \em interface for concrete flux drivers,
the \em XSecAlgorithmI \em interface for concrete cross section models, and
the \em EventRecordVisitorI \em interface for for concrete event generation steps.
Invoking all algorithms through such standardized interfaces guarantees 
scalability and ensures the seamless integration of new concrete implementations.

The algorithmic objects are stateless and their behavior is fully externally configured. 
The algorithm configurations are stored in XML files. Typically, there is a single XML 
configuration file per algorithm. Each file may contain multiple configuration sets 
for that algorithm with each configuration set being uniquely identified by a name. 
The algorithm configuration variables can be of many different types
(including booleans, integers, real numbers, strings, ROOT 1-D or 2-D histograms, ROOT n-tuples/trees 
or other GENIE algorithms with their own configurations).
Each configuration variable, in a given set, is uniquely identified by a name.
During the initialization phase, all XML configuration files are parsed and each named
configuration set is stored at a type-safe `value' $\rightarrow$ `type' associative container called the 
\em Registry \em.
All \em Registry \em objects instantiated in initialization phase are stored in a shared pool
called the {\em AlgConfigPool}. A unique name is being used to identify each {\em Registry} in that pool.
The name is constructed  by the name of the configuration set, the name of the algorithm the configuration is intended for 
and the namespace that the algorithm lives in, as `namespace::algorithm-name/configuration-name'.
At run-time each algorithmic object can look up its configuration set by accessing the corresponding \em Registry \em object.

One feature of the GENIE configuration system is especially worth noting. 
Algorithm configuration sets may include other algorithms (with their own configurations, which
in turn may contain more algorithms).   GENIE's extensibility and flexibility is largely due
to this feature in conjunction with the standardization of the algorithm interfaces.
In the actual GENIE code one only needs define a call sequence between abstract algorithm-types
such as, for example, that an algorithm-type specialized in generating scattering kinematics,
invokes another algorithm-type specialized in cross section calculations which, in its turn,
should invoke another algorithm type specialized in form factor calculations. Once that call
sequence has been defined in the code, many concrete realizations may come into being purely at 
the configuration level 
by specifying the names of the concrete algorithms and the names of their configuration sets. 

Typically, pre-configured instances of GENIE algorithms are accessed through an algorithm 
factory \cite{DesignPatterns:GangOfFour} which is responsible for instantiating each
algorithm (upon request) and allowing it to look up its configuration. The factory
typically owns and manages the list of all instantiated concrete algorithms.
Since algorithms are stateless objects, further requests for an instantiated
concrete algorithm results in the previously instantiated algorithm being returned
rather than a new one being created.

By default all instantiated concrete algorithms and configurations are stored within shared pools
designed as singletons \cite{DesignPatterns:GangOfFour}.
As these are shared pools, modifications have global effects. For example, modifying 
a low-level algorithm configuration modifies all call sequences that include that algorithm.
This is desirable in most contexts, such as for example for the consistent propagation of physics
parameter changes throughout GENIE. 
There are certain situations, however, such as fitting or event reweighting applications, where this may 
be not be a desirable feature. The GENIE Core Framework allows algorithms to clone and assume ownership 
of the entire sequence of sub-algorithms they depend upon, along with each sub-algorithm's configuration 
registries. That cloned call-sequence of algorithmic objects is stored in a local rather than a shared pool. 
In this way, concrete top-level algorithms behave as self-contained capsules and can be re-configured
in isolation without affecting other GENIE components.

%
%
%

%
%


\subsection{Event generation framework: Particles, Events and Interactions}

In this section three key framework classes, the \em GHepParticle\em,
\em GHepRecord\em, and the \em Interaction \em classes, are described. 

GENIE is using the natural system of units $(\hbar=c=1)$ so almost every simulated quantity is expressed
in powers of [GeV]. Exceptions are the event vertex in the detector coordinate system (in SI units) and
particle positions in the hit nucleus coordinate system (in fm).
Different units may be employed when native GENIE event descriptions are converted to
experiment-specific formats in accordance with the format specification. 

\subsubsection{Particles}

The basic output unit of the event generation process is a `particle'. 
This is a term used to describe both particles and nuclei
appearing in the initial, intermediate or final state, 
as well as  generator-specific pseudo-particles used for facilitating book-keeping 
of the generator actions.

Each such `particle' generated by GENIE is an instance of the \em GHepParticle \em class.
These objects contain information with particle-scope including: 
particle ID and status codes, PDG mass, charge, name, 
indices of mother and daughter particles marking possible associations with other particles in the same event,
4-momentum,  4-position in the target nucleus coordinate system, 
polarization vector, 
and other properties.
The \em GHepParticle \em class includes methods for setting and querying these properties.

GENIE has adopted the standard PDG particle codes \cite{Yao:2006px}.
For ions it has adopted a PDG extension, using the 10-digit code 10LZZZAAAI 
where 
L is the number of strange quarks 
ZZZ is the total charge,
AAA is the total baryon number, 
and I is the isomer number 
(I=0 corresponds to the ground state).
GENIE-specific pseudo-particles have PDG code $>=$ 2000000000 and can convey important
information about the underlying physics model.
Pseudo-particles generated by other specialized programs that may be called by GENIE
(such as PYTHIA-6) are allowed to retain the codes specified by that program.

GENIE obtains particle data (including particle names, codes, masses, widths, decay
channels and more) using the ROOT's  {\it TDatabasePDG}. This particle data-base manager
object is  initialized with the constants used in PYTHIA-6. The data-base has been augmented
by the GENIE authors to include baryon resonances, nuclei and GENIE-specific pseudo-particles. 
Details are given in Ref. \cite {genie-doc-753}.

GENIE marks each particle with a status code.
This signifies the position of a particle in a time-ordering 
of the event and helps navigation within the event record.
Most generated particles are marked as one of the following:
\begin{itemize}
\item `initial state', typically the first two particles of the event record corresponding to the incoming neutrino and the nuclear target.
\item `nucleon target', corresponding to the hit nucleon (if any) within the nuclear target.
\item `intermediate state', typically referring to the remnant nucleus, fragmentation intermediates such as quarks, diquarks,
      or intermediate pseudo-particles.
\item `hadron in the nucleus', referring to a particle of the primary hadronic system,
      defined as the particles emerging from the primary interaction vertex before any possible
      re-interactions in the nucleus.
\item `decayed state', such as for example unstable particles that have been decayed.
\item `stable final state' for the relatively long-lived particles emerging from the nuclear targets.
\end{itemize}

All particles generated by GENIE during the simulation of a single neutrino interaction
are stored in a dynamic container representing an `event'.

\subsubsection{Events}

Events generated by GENIE are stored in a custom, STDHEP-like event record called a \em GHEP \em record.
Each \em GHEP \em event record, an instance of the \em GHepRecord \em class, 
is a ROOT TClonesArray container of \em GHepParticle \em objects representing individual particles.

Other than being a container for the generated particles,
the event record holds additional information with event-, rather than particle-, scope
such as the cross sections for the selected event, 
the differential cross section for the selected event kinematics, 
the event weight, 
a series of customizable event flags, 
and interaction summary information (described in the next section).

Additionally, the event record includes a host of methods for querying / setting event properties
including many methods that allow querying for specific particles within the event.
Examples include methods to return the target nucleus, the final state primary lepton, 
or a list of all stable descendants of any intermediate particle.

The event record features a `spontaneous re-arrangement' feature
which maintains the compactness of the daughter lists at any given time.
This is necessary for the correct interpretation of the stored particle associations as 
the daughter indices correspond to a contiguous range.
The particle mother and daughter indices for all particles in the event record 
are automatically updated as a result of any such spontaneous particle rearrangement.

The event generation itself is built around the \em GHEP \em event record using the
Visitor design pattern \cite{DesignPatterns:GangOfFour}. 
The interaction between the \em GHEP \em event record and the event generation
code will be outlined in the following sections.

The \em GHEP \em structure is highly compatible with the event structures used in most HEP generators.
That allows us to call other generators, such as PYTHIA-6, 
as part of an event generation chain and convert / append their output into the current \em GHEP \em event.
Additionally the \em GHEP \em events can be converted to many other formats for facilitating 
the GENIE interface with experiment-specific offline software systems and cross-generator comparisons.

\subsubsection{Interactions}

The \em GHEP \em record represents the most complete description of a generated event.
Certain external heavy-weight applications such as 
specialized event reweighting schemes or
realistic detector simulation chains using the generator as the physics front-end 
require all of the detailed particle-level information.
However, many of the actual physics models employed by the generator, 
such as cross section, form factor, or structure function models, require a much smaller subset of 
information about the event.  

An event description based on simple summary information,
typically including a description of the initial state, the process type and the scattering kinematics,
is sufficient for driving the algorithmic objects implementing these physics models.
In the interest of decoupling the physics models 
from event generation and the particle-level event description,
GENIE uses an \em Interaction \em object to store summary event information.
Whenever possible, algorithmic objects implementing physics models 
accept a single \em Interaction \em object as their sole source of information about an event.
This enables the use of these models 
both within the event generation framework 
but also within a host of external applications such as
model validation frameworks, event re-weighting tools and user physics analysis code.

An \em Interaction \em object is an aggregate, hierarchical structure, 
containing many specialized objects holding 
information for the initial state (\em InitialState \em object), 
the event kinematics (\em Kinematics \em object), 
the process type (\em ProcessInfo \em object) and 
potential additional information for tagging exclusive channels (\em XclsTag \em object).
Instantiating \em Interaction \em objects for driving physics models
is streamlined using the `named constructor' C++ idiom. 
They can be serialized into unique string codes which,
within the GENIE framework, play the role of the `reaction codes' of the old procedural systems.
These string codes are used extensively for mapping information to interaction types.
Two examples include mapping interaction types to pre-computed cross section splines 
or mapping interaction types to specialized event generation code.
Each generated event has an \em Interaction \em summary object already attached to it.

\subsection{Event generation processing units: Modules, Threads and Drivers}

On an operational level the responsibility for generating events is shared
between event generation \em drivers, threads \em and \em modules. \em  Tasks are delegated 
from event generation drivers to threads, and from threads to modules.   
Event generation drivers can include multiple threads, and threads can include multiple modules.
Event generation drivers are responsible for generating events for a particular user-defined
situation.  These can be as simple as monoenergetic neutrinos interacting off a single target, 
to complex situations involving the output of realistic beam-line simulations and full 
detector geometry descriptions.  
Threads are responsible for generating the physics of particular classes of events, for instance 
charged-current quasielastic.   Modules carry out a single step in that generation process.


\subsubsection{Event generation modules}

An event generation module is a key event generation abstraction. 
Each event generation module encapsulates a well defined event generator 
operation which, in physics terms, can be any of a very diverse set of actions. 
Examples include
selecting the scattering kinematics, generating the final state primary lepton 
or the primary hadronic system, 
transporting hadrons within the target nucleus, and decaying unstable particles.

Operationally, event generation can be seen as a series of well-defined processing 
steps operating on the \em GHEP \em event record.  
The act of operating on the 
event record defines an interface that is encapsulated by the \em EventRecordVisitorI \em
abstract class. 
As it is indicated by the interface name, 
the Visitor design pattern is being employed \cite{DesignPatterns:GangOfFour}.   
Concrete event generation modules,
implementing the \em EventRecordVisitorI \em interface,  
`visit' the event record. 
The event record then invokes each attached module and is modified as a result.

Due to the diversity of the event processing operations that must  be considered by GENIE, 
we formed the event generation module abstraction focusing on the common operational 
aspect of (potentially) modifying the event record. 
This represents the most generic way
of thinking about event generation and guarantees that any future physics addition,
especially ones not envisioned at this stage of the generation evolution, 
can be trivially embedded into the existing framework.
Treating the event generator modules uniformly and standardizing on the event generation module interface 
allows us to build a flexible and extensible system where modules can be dynamically 
plugged in/out of the event generation or interchanged. 
Examples can further clarify the utility of this abstraction:
 a module handling a set of particle decays can be unplugged 
to inhibit those decays, or a module handling intra-nuclear hadron transport may be swapped with another module 
performing the same operation using a different physics strategy.

Whenever possible event generation modules are written in a generic way,
containing code implementing just the neutrino event generation mechanics.
The actual physics model itself is specified in the generation module configuration.
This decoupling of mechanics from models greatly simplifies code development, transparency, 
and physics validation, simplifying the overall structure and 
reducing the amount of code that needs to be actively developed and scrutinized between successive releases. 
An example will clarify this factorization:
The module selecting the kinematics for deep-inelastic neutrino-nucleon interactions 
does not contain the actual code for the deep-inelastic differential cross section. It merely contains code to
calculate the allowed kinematical phase space for the process, select a point in that phase space using a 
Monte Carlo acceptance / rejection method, and update the \em GHEP \em record accordingly. 
The actual differential cross section model used during the Monte Carlo selection 
is an external physics model invoked by the event generation module. The module itself can
be recycled many times by instructing it to call a different cross section model each time. 
As a result of that factorization, multiple call sequences can be defined purely at the configuration level
without code duplication. 

\subsubsection{Event generation threads}

An \em event generation thread \em is an ordered sequence of 
processing steps, encapsulated by event generation modules, that can  be applied to an 
empty \em GHEP \em event record to completely generate some class of physics events.
This process defines an 
interface that is encapsulated by the \em EventGeneratorI \em abstract class.
Within the GENIE event generation framework the structures containing a comprehensive set of 
instructions for generating a class of physics events are concrete \em EventGeneratorI \em objects. 

GENIE defines a comprehensive set of event generation threads responsible for generating 
event types at the level of fundamental interactions.   
The complete set of these event generation threads comprises GENIE's 
full `physics content' for event generation.   
As an event generation thread can generate a single class of events only, there are 
usually multiple threads in use. 

The class of physics events generated by a thread 
can have an arbitrary granularity, from a single interaction
corresponding to a particular process type with a given final state 
to very broad event categories.
Each thread contains an \em InteractionList \em object, a container 
containing a list of 
the \em Interaction \em objects the thread can generate. 
The \em InteractionList \em plays a crucial role in 
identifying the responsibilities of each thread within the GENIE framework. 
Once an event type to be generated has been selected, a corresponding \em Interaction \em object
is instantiated. Following the Chain of Responsibility design pattern \cite{DesignPatterns:GangOfFour}, 
GENIE attempts to match the 
\em Interaction \em object with an element of the \em InteractionList \em containers for all active threads. 
The first thread found that is able to handle that event type is handed the responsibility
to generate the event. 

Additionally, event generation threads include an instance of the cross section algorithm
that can be used for selecting the event kinematics or for computing the probability for
a particular neutrino to interact.  
This is another example of separating mechanics from models
and serves to greatly simplify the dynamic mapping between event types and cross section
models.  

Once a list of threads has been loaded into the generator, many high-level event generation
operations became trivial.   Compiling the list of all event types that can be generated
by GENIE in its current configuration simply involves looping over the active threads
and adding the corresponding \em InteractionList \em objects. 
Selecting an event type to be generated from that master list involves looping over 
its \em Interaction \em objects and, for each element, identifying the responsible thread, 
requesting its corresponding cross section model and invoking it by 
passing the \em Interaction \em object as argument.
Once an event type has been selecting generating the event simply involves looking up
the responsible thread and delegating responsibility to it.

During event generation an invoked thread maintains a modification history of the event record.
If a tried 
event generation path leads to a dead-end, the current event generation module throws an exception and aborts. 
The event generation thread catches that exception and, depending on information stored at it, may rerun the event 
using a snapshot of the event record taken N steps back, in the hopes of taking an 
alternative path and avoiding the encountered dead-end. 
If a configurable maximum number of exceptions is caught, or if any thrown exception specifies explicitly that 
generation of the current event is to be aborted altogether, the thread sets the appropriate error flags and makes 
sure that the remaining processing steps are skipped.
The user, via options set in the event generation driver, may choose to keep certain types of these events so as
to examine their type and frequency, though the default behavior of GENIE is
to discard these events and only write out physical, fully formed events.
Error handling within each active thread greatly adds to the robustness and fault-tolerance
of the package, which is especially valued in large-scale, CPU-intensive, experiment-specific Monte Carlo 
production runs involving hundreds of CPU cores over many weeks.

Advanced users can modify the default event generation threads by removing / adding event generation modules, 
or they can define their own uniquely named threads for handling new processes or handling existing processes in new ways.

\subsubsection{Event generation driver classes}

GENIE provides two event generation driver classes.  
These drivers collect the user inputs, instantiate and configure 
all required event generation components,
and oversee communications 
between these components, the computing environment, and
the user.

The two driver classes support two different types of functionality:
\begin{itemize}
\item 
Instances of the \em GEVGDriver \em class can handle event generation for a 
given initial state corresponding to an arbitrary neutrino / target pair.
\item 
Instances of the \em GMCJDriver \em class can be used for more complicated
simulations involving arbitrarily complex, realistic beam flux simulations
and detector geometry descriptions. 
This driver object concerns itself mostly with driving the flux and detector 
geometry navigation drivers and integrating those with the GENIE event generation framework.
It represents a significantly more complex and CPU-intensive event generation case 
but relies entirely on a host of \em GEVGDriver \em instantiations, one for each 
possible initial state in that Monte Carlo job, 
in order to obtain neutrino interaction physics modeling capabilities and
generate event kinematics.
\end{itemize}


\begin{figure*}[htb]
\center
\includegraphics[width=32pc]{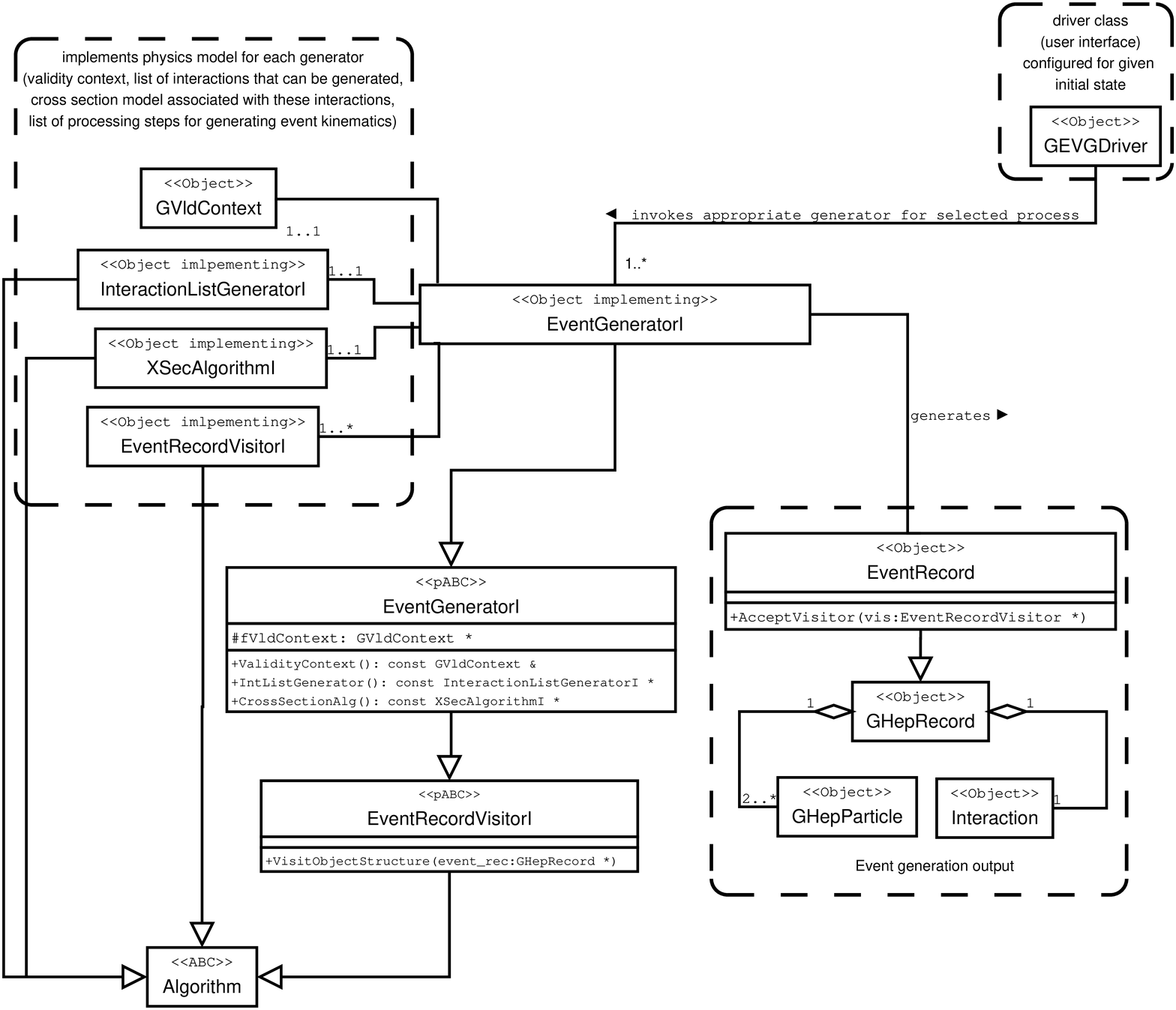}
\caption{A UML diagram depicting the GENIE event generation framework. See text for details.}
\label{fig:fmw}
\end{figure*}

\section{GENIE Event Generation Applications and Utilities} 
\label{sec:apps}

GENIE is being used by a host of precision-era neutrino experiments and 
provides off-the-shelf components for generating neutrino interactions under the
most realistic assumptions. 
The event generation driver classes described in Section \ref{sec:design}
are encapsulated within driver
applications which present the user with a command-line or graphical interface,
instantiate and configure those driver classes, call the event generation methods to
generate the requested number of events, and push those events through a persistency
manager. 

In experiment-specific GENIE-based event generation drivers utilizing the 
\em GMCJDriver \em one can integrate 
the GENIE neutrino interaction modeling with detailed flux and detector geometry descriptions.
This is a non-trivial operational capability that older procedural neutrino generators 
typically lacked, requiring significant development effort from experiments.
The flux descriptions are typically derived from experiment-specific beam-line simulations
while the detector geometry descriptions are typically derived from engineering drawings
mapped into the GEANT4 \cite{Agostinelli:2002hh}, ROOT \cite{Brun:1997pa} or 
GDML \cite{Chytracek:2006be} geometry description languages.
Obviously, flux and detector geometry descriptions can take many forms, driven by
experiment-specific choices. 
GENIE standardizes the geometry and flux driver interfaces by defining the operations
that GENIE needs to perform on the geometry and flux descriptions and the essential flux and
geometry information needed for the generation of events. 
Concrete implementations of the interfaces allow experiments to extend GENIE's event
generation capabilities and make it possible to seamlessly integrate new geometry 
descriptions and beam fluxes into user applications.  

In this section we will describe in some detail the flux and geometry interfaces.  
We will briefly describe applications built from these drivers as well as 
GENIE utilities to evaluate and display cross section information, make comparisons 
to external data, and facilitate model tuning.  


\subsection{Neutrino flux drivers}

In GENIE every concrete flux driver implements the \em GFluxI \em interface.
The interface defines what neutrino flux information is needed by the event 
generation drivers and how that information is to be obtained.
Each concrete flux driver includes methods to:

\begin{itemize}
\item Declare the neutrino flavors that can generate events. 
      This information is used for initialization purposes, in order to construct a list
      of all possible initial states in a given event generation run.
\item Declare the maximum energy. 
      Again this information is used for initialization purposes, in order to calculate
      the maximum possible interaction probability in a given event generation run. 
      Since neutrino interaction probabilities are tiny,
      GENIE scales all interaction probabilities
      in a particular event generation run so that the maximum possible interaction probability is 1.
      That maximum interaction probability corresponds to the total interaction probability (summed
      over nuclear targets and process types) for a maximum energy neutrino following a trajectory
      that maximizes the density-weighted path-lengths for each nuclear target in the geometry.
      GENIE adjusts the MC run normalization accordingly to account for this probability renormalization.
\item Generate a flux neutrino and specify its pdg code, its weight (if any),
      its 4-momentum and 4-position. 
      The 4-position is given in the detector coordinate system (as specified by the input geometry).
      Each such flux neutrino is propagated towards the detector geometry but is not required to
      cross any detector volume. GENIE will take that neutrino through the geometry, 
      calculate density-weighted path-lengths for all nuclear targets in the geometry,
      calculate the corresponding probabilty for scattering off each nuclear target and 
      decide whether that flux neutrino should interact. If it interacts, an appropriate 
      \em GEVGDriver \em will be invoked to generate the event.
\item Notify that no more flux neutrinos can be thrown. 
      Flux drivers that use the output of beam-line simulations, so-called `flux files', 
      are configured to recycle these flux files multiple times in a given run since 
      most neutrino flux entries do not produce an interaction. 
      The flag allows GENIE to 
      properly terminate the event generation run once this limit is reached, irrespective
      of the accumulated number of events, protons on target, or other metric of exposure.
\end{itemize}

The above correspond to the common set of operations/information that GENIE expects to
be able to perform/extract from all concrete flux drivers.
Specialized drivers may define additional information that can be utilized in experiment-specific
applications.   One typical example of this is to `pass-through' 
information about the flux neutrino parents placed in the flux files by the beamline simulation, 
such as the parent meson PDG code, its 4-momentum, and its 4-position at the production and decay points.  

At the time of writing this article, GENIE already includes a host of concrete 
flux drivers allowing GENIE to be used in many realistic, experiment-specific situations.
More specifically, it includes an interface to the JPARC neutrino beam simulation \cite{JNUBEAM:PrivCom}
used by Super-Kamiokande \cite{Fukuda:2002uc}, nd280 \cite{Wark:2008zz}, and INGRID \cite{Wark:2008zz}
and an interface to the NuMI beam simulation \cite{Anderson:1998zz}
used by MINOS \cite{Michael:2008bc}, NOvA \cite{Ayres:2004js}, MINERvA \cite{Schulte:2006db}, 
MicroBooNE \cite{Soderberg:2009rz} and ArgoNEUT \cite{Soderberg:2009qt}. 
It includes drivers for the BGLRS \cite{Barr:2004br} and the FLUKA \cite{Battistoni:2001sw} atmospheric fluxes.
Moreover, it includes a  generic flux driver,
describing a cylindrical neutrino flux of arbitrary 3-D direction and radius,
with a configurable radial dependence,
which can be used for describing a flux containing a number of different neutrino species whose 
(relatively normalized) energy spectra are described as 1-D histograms.
GENIE, obviously, also includes the trivial case of a monoenergetic flux typically employed 
in physics benchmarking calculations.

Concluding this section
it is worth re-emphasizing the fact that new concrete flux drivers (describing the neutrino flux
from other beam-lines) can be easily developed and they can be effortlessly and seamlessly 
integrated within the GENIE event generation framework.

\subsection{Geometry drivers}

In GENIE every concrete geometry driver implements the \em GeomAnalyzerI \em interface.
The interface specifies what information about the input geometry is relevant to the
event generation and how that information is to be obtained. 
Each concrete geometry driver implements methods to:

\begin{itemize}
\item Declare the list of target nuclei that can be found in the geometry. 
      This information is used for initialization purposes, in order to construct a list
      of all possible initial states in a given event generation run.
\item Compute the maximum density-weighted path-lengths for each nuclear target in the geometry.
      Again, this is information used for initialization purposes. The computed `worst-case'
      trajectory is used to calculate the maximum possible interaction probability in a particular
      event generation run. This maximum interaction probability is used internally to normalize 
      all computed interaction probabilities in that run.
\item Compute density-weighted path-lengths for all nuclear targets, for a trajectory of given 4-momentum and starting 4-position.
      This allows GENIE to calculate probabilities for each flux neutrino to be scattered off every nuclear target
      along its path through the detector geometry.      
\item Generate a vertex along a trajectory of given 4-momentum and starting 4-position on a volume containing a given nuclear target.
      This allows GENIE to place a neutrino interaction vertex within the detector geometry once an interaction of a flux
      neutrino off a selected nuclear target has been generated.
\end{itemize}

GENIE currently contains a concrete geometry driver able to handle the ROOT-based detector geometry descriptions
typically used by most neutrino experiments. Detector geometry descriptions based on GEANT or GDML can be 
converted into ROOT-based descriptions and used by the same GENIE geometry driver as well.
GENIE also includes a driver for more trivial geometry descriptions corresponding to a single nuclear 
target or a target mix (a set of nuclear targets, each with its corresponding weight fraction) at a fixed 
position. This simpler geometry driver may be used in simulating fixed initial states for benchmarking calculations 
or in experimental situations where a relatively  uniform detector is being illuminated by a spatially 
uniform neutrino beam.  An example of the latter would be a detector placed far enough 
from the beam-line instrumentation so as to see a point-like neutrino source.

Again it is worth re-emphasizing that any new detector geometry description can be seamlessly 
integrated with the GENIE event generation framework by means of developing an appropriate GENIE geometry driver.

\subsection{Event generation outputs}

The generated events are stored in the ROOT file format. The typical output of an event generation run is a single 
ROOT file which contains an event tree with a single branch and a single leaf per event containing the generated 
\em GHEP \em record.
User-defined branches to write out experiment-specific information
may be added to that tree with the user-defined information `linked' to the corresponding generated neutrino event.
The output ROOT file contains directories storing all GENIE configuration information for the MC run and a snapshot
of the running environment for later reference.
GENIE provides a persistency manager object which can be employed within the event generation driver applications 
to write out the event tree.

\subsection{Event generation applications}

GENIE is distributed with many event generation applications and utilities, 
many of which are straightforward wrappers for the 
components described above.  
Users interact with these applications through simple command-line interfaces, 
user-created XML configuration files, and environmental variables.   

The \em gevgen \em application can be used in simulating given initial states 
for benchmarking calculations or simple
experimental setups for which histogram-based flux descriptions and simple geometry descriptions in terms
of a target mix are adequate.   Experiment-specific event generation applications, 
such as \em gT2Kevgen \em and \em gNuMIevgen\em,
employ the detailed JPARC and NuMI beam-line simulations and the ROOT-based detector geometry descriptions of
the corresponding experiments and are used by a large fraction of the GENIE user base.

On a MacBookPro running MAC OS X 10.5.6, with a 2.16 GHz Intel Core 2 Duo and
1 GB DDR2 SDRAM at 667 MHz, and with all event generation threads enabled,
GENIE simulates around 70 events/sec for a $\nu_{\mu}+Fe^{56}$ initial state at $E_{\nu} = 1 GeV$.
The speed is 5 events/sec with the detailed nd280 ROOT-based detector geometry description
  (40 nuclear targets) and the detailed JNUBEAM-based JPARC neutrino beam simulation
  (4 neutrino species). 

\subsection{Utilities}

Event generation for a realistic experimental setup typically requires of the order of $\sim 10^{8} - 10^{9}$ differential
cross section evaluations just in order to select an interaction to be generated. 
It is therefore very practical to perform the numerical 
differential cross section integrations at a different stage and save the data for building cross section splines. 
The event generation components can recycle these splines for performing fast numerical interpolations,
greatly improving the event generation efficiency.
GENIE provides a utility, \em gmkspl\em, to generate all required cross section splines for the intended set of neutrino
flavors and nuclear targets over the required energy range and write the data in the XML format expected by the event
generation components.

For many use-cases it is convenient to analyze the output \em GHEP \em ROOT event tree and either write-out simpler, flat
n-ntuples containing summary information or convert it to a format expected by an experiment-specific application, such as
a detector-level simulation that doesn't use the GENIE I/O.
GENIE provides an event tree converter, \em gntpc \em, which writes out a host of alternative plain text, XML or bare-ROOT
formats currently is use by GENIE-based applications in client experiments.

Several tools exist for the purposes of validating and tuning the physics models in GENIE.   
The ultimate source of data for many of these comparisons is the DURHAM Neutrino Scattering 
Data Resource \cite{Whalley:2004sz}, an online resource with large compilations of neutrino data from 
many different experiments.    
Access to this data in the GENIE package is done through the NuValidator \cite{Andreopoulos:2005vd} 
program, a GENIE add-on that can 
be optionally installed.  Distributed with the NuValidator in XML format are 
data from the DURHAM database, electron 
scattering data from the Jefferson Lab database \cite{e49a10}, and publicly available 
lepton scattering structure function data \cite{Gehrmann:1999xn}.
Electron scattering data and lepton structure function 
data are important in evaluating any scheme for handling the perturbative / non-perturbative 
transition region described in Sec. \ref{sec:transition} \cite{Gallagher:2004nq}.
This data is then available for a variety of applications.  The NuValidator includes a simple GUI interface
allowing one to select data to display together with the GENIE prediction, with full ability to set model 
parameters to new values.  
The data can also be accessed for physics model parameter fits 
possibly including new experimental data as well as external, historical data.      

In addition to event generation and validation the flexibility of the GENIE framework 
also simplifies many downstream analysis tasks.  The evaluation of generator-related
systematic errors can be greatly simplified through event reweighting 
\cite{dobson-ladek}. 
As mentioned in Sec. \ref{sec:core}, 
GENIE's ability to allow algorithm configuration using local pools, redundancy of 
information between \em GHEP \em event records and \em Interaction \em objects,
and ability to identify algorithms/configurations in generated output 
all support the development of experiment-specific event reweighting programs. 

Details on the application described above and on a host of other utilities 
may be found in the GENIE Physics and User Manual \cite{genie-doc-753}.

\section{Collaboration Organization}
\label{sec:collab}

A significant organizational challenge for the GENIE project is defining the
working relationship with experimental collaborations.  
Having many experiments use the same neutrino 
event generator is a new development in this field  and the exact nature of these
working relationships will evolve over time.  
However some realities and goals are clear.   

Good two-way communication is essential both for immediate issues like 
bug reporting and support and for longer term issues like planning of upcoming 
physics releases.   Experiments often set target dates to begin production 
of Monte Carlo samples based on publication plans and expectations for data-taking. 
Meeting these deadlines is often a high priority for these collaborations.  Since the 
time-scale for production of large Monte Carlo samples is roughly similar to the 
timescale for production of new GENIE physics releases, it will be desirable to 
synchronize, or at least be cognizant of, upcoming experimental deadlines to 
as large a degree possible.  Discussions with collaborations will also focus
on priorities for physics model improvements.  

One conduit for these collaborations will be through experimental liaisons who serve 
as the main contact within an experiment on GENIE issues.  
This person can then report on the experiment's experiences and deadlines, 
and can help present the experiment's priorities for model improvements to the 
GENIE collaboration for evaluation.  When effort within the GENIE collaboration to 
incorporate new model work is not available, these liaisons can assist in providing
and organizing effort from their collaborations.       

Physics model development is partitioned into subtasks including the cross section model,
the hadronization model, and the intranuclear rescattering model. These components
are all relatively self-contained and have validation procedures independent of the others. 
Overall tuning and validation of a production release is the responsibility of a separate 
working group of the collaboration.  This task is undertaken on a roughly yearly timescale.  
This exercise finalizes the model set and determines values of all parameters based on 
fits to external data.  
This process also determines parameter errors which can then be used by experiments in the 
evaluation of generator-related systematic errors. 

A default tune - a self consistent set of physics models and parameter values - 
will be specified by the collaboration for every major release, along with 
information about the uncertainties on model parameters and possible correlations.    
It is possible for experiments, using the validation and parameter fitting tools
that are provided as part of the GENIE package, 
to have their own `tuning' of the generator.  This would be desirable from 
an experiment's perspective as they have access to new data that, due to the aforementioned 
uncertainties, will probably not be in complete agreement with the default tuning.   
It is hoped that the development of new models by specific collaborations as part of 
ongoing analyses will be speedily adopted 
into GENIE for the benefit of other users, though the decision 
of when to make their new models public will be left to the discretion of the 
user collaborations.   It will, however,  
be important that model improvements be merged back 
into the default tuning on a regular basis so as to prevent the fragmentation of the
base set of physics models into many different experiment-tuned versions which then 
evolve independently over many years.  

The collaboration organizes its effort internally through occasionall meetings
of physics model and tuning/validation working groups, phone meetings, blogs, 
and web-based document databases.  
The GENIE web site \cite{GENIE} is the central repository 
for all information related to the package. 
An extensive Physics and User Manual \cite{genie-doc-753}, a web-based source
code Reference Manual \cite{GENIE:Doxygen} as well as a support mailing list 
are available to GENIE users.
Hands-on tutorials on GENIE have been given on several occasions at
meetings in the U.S., Europe, and Japan, and material from these workshops with
introductory and advanced tutorial examples are available on the GENIE web site.

%
%

\section{Code Availability}
\label{sec:avail}

\subsection{Version control and distribution}

GENIE is available from its CVS repository hosted at STFC's Rutherford Appleton Laboratory
from where one can access the development version and a series of `frozen' production-quality releases.
The repository is physically located on AFS space and, in read-only mode, 
can be accessed anonymously. Read-write mode access to the code repository 
is provided to the GENIE collaborators via SSH and public key authentication.
A transition to a SubVersion repository is planned for the near future.
Up-to-date details on how to access the source code are given at the
GENIE web-site \cite{GENIE} and at the Physics and User Manual \cite{genie-doc-753}.

\subsection{Supported platforms and external dependencies}

GENIE is known to build on many platforms, including all popular LINUX 
distributions and MAC OS X,
and has no OS proprietary dependency. 
As of GENIE v2.5.1, external dependencies for a minimal installation that can
be used for physics MC production include the
\em ROOT Class Libraries \em \cite{Brun:1997pa}, 
the \em LHAPDF parton density function library \em \cite{Whalley:2005nh}, 
the \em PYTHIA-6 \em LUND MC \cite{Sjostrand:2006za}
and two fairly common utilities: 
the \em libxml2 \em XML parser and the \em log4cpp \em error logger. 

\subsection{Versioning scheme and release lifetime}
 
GENIE versions are numbered as `$i.j.k$', where $i$, $j$ and $k$ are the
major, minor and revision indices respectively. The corresponding CVS tag is `$R-i\_j\_k$'.
When a number of significant functionality improvements or additions have been made, 
the major index is incremented. 
The minor index is incremented in case of significant fixes or minor feature additions.
The revision number is incremented for minor bug fixes and updates.
Versions with even minor number correspond to validated, physics production releases.
Versions with odd minor number correspond to the development version of `candidate' 
releases tagged during the validation stage preceding a physics production release.
Tagged versions always have an even revision number. Odd revision numbers correspond to 
the CVS head.

Because of the effort invested by the experimental communities in generating large 
samples and understanding the impact of the simulation changes into their physics results,
physics production releases are nominally supported for a long term of approximately 18 - 24 months.

\subsection{License}

GENIE is now distributed under the GPLv3 license agreement \cite{GPLv3}.

%
%

\section{Conclusions}

GENIE provides a modern and versatile platform for 
a universal, `canonical' Neutrino Interaction Physics Monte Carlo
whose validity will extend to all nuclear targets and neutrino flavors over a
wide range of energies from MeV to PeV scales.
Currently, it includes state-of-the-art neutrino interaction physics modeling in the few-GeV
energy range which is relevant for the current and near future
long-baseline precision neutrino experiments using accelerator-made beams.  

The software was designed using object-oriented methodologies and developed entirely in 
C++ over a period of more than three years, from 2004 to 2007.  
The design of the package decouples the mechanics of event generation from 
the physics models, providing modularity, extensibility, and flexibility.  The package supports
the full life-cycle of generator-related activities, 
from event generation using detailed detector geometries and flux information,
to final analysis tasks such as reweighting and systematic error evaluation.   
The data, programs, and procedures used to validate and tune the package are all distributed with 
the package itself, allowing users the ability to easily extend the package and evaluate new
models.   

The project is supported by a group of physicists from all major experiments
operating in this energy range, establishing GENIE as a major HEP event generator collaboration.
GENIE has already been adopted by many neutrino experiments, including
those using the JPARC and NuMI neutrino beamlines, and will be an important physics tool   
for the worldwide accelerator neutrino program.

\section*{Acknowledgements}

This work was supported by the UK Science and Technology Facilities Council  / Rutherford Appleton Laboratory, 
the US Department of Energy, the US National Science Foundation, and the Tufts University Summer Scholars Program.  

The authors would like to express our gratitude to 
G.Irwin (Stanford), 
B.Viren (Brookhaven Lab), 
S.Kasahara (Minnesota) and 
N.West (Oxford) 
for contributing to the early stages of the evolution of GENIE
through example, by developing the MINOS offline framework, and
through their inputs and criticisms during the GENIE design reviews.

We would also like to thank our MINOS collaborators, in particular
R.Gran (UMD), 
K.Hofmann (Tufts), 
M.Kim (Pittsburgh), 
M.Kordosky (UCL), 
W.A.Mann (Tufts), 
J.Morfin (Fermilab) and
S.Wojcicki (Stanford),
for their contributions in the development, tuning and validation of
the default set of physics models in GENIE.

We would also like to thank 
E. Paschos (Dortmund), 
S. Mashnik (Los Alamos),
A.Bodek (Rochester),
O.Lalakulich (Dortmund, Giessen) and 
T.Leitner (Giessen)
for providing important models and results.
We also express our gratitude and recognition to
C.Reed (NIKHEF), 
K.Scholberg, C.Little, R.Wendell (Duke), 
A.Habig, R.Schmidt (UMN Duluth) and 
D.Markoff (NCCU) 
for their ongoing effort 
to extend the GENIE validity range down to the MeV energy scale.

Also we would like to thank  
C.Backhouse (Oxford), 
S.Boyd (Warwick), 
J.J. Gomez Cadenas (IFIC Valencia),
Y.Hayato (ICRR),
J.Holeczek (Silesia),
Z.Krahn (Minnesota),
H.Lee (Rochester),
J.Lagoda (L'Aquila),
S.Manly (Rochester),
B.Morgan (Warwick), 
D.Orme (Imperial),
G.Perdue (Fermilab),
T.Raufer (RAL),
D.Schmitz (Fermilab),
E.Schulte (Rutgers), 
J.Sobczyk (Wroclaw),
A.Sousa (Oxford), 
J.Spitz (Yale),
P.Stamoulis (Athens), 
R.Tacik (Regina),
H.Tanaka (UBC),
I.Taylor (Imperial), 
R.Terri (QMUL), 
V.Tvaskis (Victoria),
Y.Uchida (Imperial), 
S.Wood (JLAB), 
S.Zeller (LANL), 
L.Zhu (Hampton)
and many others who have used early versions of GENIE
for their help in improving the build system, fixing bugs, and
contributing comments and tools.

%
%

\bibliographystyle{elsart-num}

\hyphenation{Post-Script Sprin-ger}

\end{document}